\documentclass[fleqn,usenatbib]{mnras}

\usepackage[T1]{fontenc}

\DeclareRobustCommand{\VAN}[3]{#2}
\let\VANthebibliography\thebibliography
\def\thebibliography{\DeclareRobustCommand{\VAN}[3]{##3}\VANthebibliography}

\usepackage{graphicx}	
\usepackage{amsmath}	
\usepackage{amssymb}	
\usepackage{newtxtext,newtxmath}

\title[EM counterparts of BNS mergers]{Host galaxies and electromagnetic counterparts to binary neutron star mergers
across the cosmic time: Detectability of GW170817-like events}

\author[Perna et al]{Rosalba Perna$^{1,2}$, M. Celeste Artale$^{3,4,5}$, Yi-Han Wang$^1$, Michela Mapelli$^{5,6,7}$, Davide Lazzati$^8$,\newauthor
 {Cecilia Sgalletta$^{5,9}$, Filippo Santoliquido$^{5,6}$}
\\
$^{1}$Department of Physics and Astronomy, Stony Brook University, Stony Brook, NY, 11794, USA\\
$^{2}$Center for Computational Astrophysics, Flatiron Institute, 162 5th Avenue, New York, NY 10010, USA\\
$^{3}$ Institut für Astro- und Teilchenphysik, Universität Innsbruck, Technikerstrasse 25/8, 6020 Innsbruck, Austria\\ 
$^{4}${Department of Physics and Astronomy, Purdue University, 525 Northwestern Avenue, West Lafayette, IN 47907, USA}\\
$^{5}$ Physics and Astronomy Department Galileo Galilei, University of Padova, Vicolo dell’Osservatorio 3, I-35122, Padova, Italy\\
$^{6}$ INFN - Padova, Via Marzolo 8, I-35131 Padova, Italy\\
$^{7}$ INAF-Osservatorio Astronomico di Padova, Vicolo dell’Osservatorio 5, I-35122, Padova, Italy\\
$^8$ Department of Physics, Oregon State University, 301 Weniger Hall, Corvallis, OR 97331, USA\\
$^9$ SISSA, via Bonomea 265, I-34136 Trieste, Italy
}

\date{Accepted XXX. Received YYY; in original form ZZZ}

\pubyear{2015}

\begin{document}
\label{firstpage}
\pagerange{\pageref{firstpage}--\pageref{lastpage}}
\maketitle

\begin{abstract}

The detection of electromagnetic radiation (EM) accompanying the
gravitational wave (GW) signal from the binary neutron star (BNS)
merger GW170817 has revealed that these systems constitute at least a
fraction of the progenitors of short gamma-ray bursts (SGRBs).  As
gravitational wave detectors keep pushing their detection horizons, it
is important to assess coupled GW/EM probabilities, and how to maximize
observational prospects. Here we perform population synthesis
calculations of BNS evolution with the code {\tt MOBSE}, and seed the
binaries in galaxies at three representative redshifts
($z=0.01,0.1,1$) of the Illustris TNG50 simulation. The binaries are
evolved and their locations numerically tracked in the host galactic
potentials until merger.  Adopting the astrophysical parameters of
GRB170817A as a prototype, we numerically compute the broadband
lightcurves of jets from BNS mergers, with the afterglow
brightness depending on the local medium density at the merger
sites. We perform Monte Carlo simulations of the resulting EM
population assuming either a random viewing angle with respect to the
jet, or a jet aligned with the orbital angular momentum of the binary,
which biases the viewing angle probability for GW-triggered events.
We find that $\sim70-80\%$ of BNSs from $z=0.01$
should be detectable in gamma-rays.  The afterglow detection
probabilities of GW-triggered BNS mergers vary between $\sim
0.3-0.7\%$, with higher values for jets aligned with the BNS angular
momentum, and are comparable across the high and
low-energy bands, unlike $\gamma$-ray-triggered events
(cosmological SGRBs) which are significantly brighter at higher
energies. We further quantify observational biases with	respect	to host	galaxy masses.

\end{abstract}




\section{Introduction}

The almost simultaneous detection of gravitational waves and broadband photons ($\gamma$-rays through radio) from a binary neutron star (BNS) merger was a historical event (The LIGO and The Virgo Collaborations, 2017). It heralded the new age of multi-messenger astrophysics, an era in which the combination of different messengers (in this case gravitational waves and photons) gives information about a source that could have not otherwise been revealed without the multi-messenger synergy
\citep{Alexander2017,Alexander2018,Hallinan2017,Haggard2017,Kasliwal2017,Troja2017,Margutti2017,Margutti2018,Lyman2018,Dobie2018,Ruan2018,Resmi2018,Piro2019,Lamb2019}.

Understanding the astrophysical origin of the binary mergers  detected by LIGO--Virgo is of much interest, and deeply connected to the
environments in which these events occur. Events in galactic fields are more likely associated with compact object (CO) binaries formed via
stellar evolution from binary stars (e.g. \citealt{tutukov1973,portegieszwart1998,belczynski2002,voss2003,belczynski2007,podsiadlowski2004,eldridge2016,marchant2016,belczynski2016,demink2016,Belczynski2017,stevenson2017,giacobbo2018,vignagomez2018,Belczynski2018,spera2019,tanikawa2020}), whereas mergers in globular clusters (e.g. \citealt{portegieszwart2000,downing2010,samsing2014,rodriguez2015,rodriguez2016,Antonini2016b,askar2017,samsing2018,rodriguez2018,fragione2019,Fragione2019clua,Fragione2019club,zevin2019,fragione2020,antonini2020,Mapelli2021a}), young star clusters (e.g. \citealt{banerjee2010,ziosi2014,mapelli2016,banerjee2017,kumamoto2019,dicarlo2019a,dicarlo2019b,Perna2019,dicarlo2020,banerjee2020,rastello2020,Kremer2020,santoliquido2020}), or disks of active galactic nuclei (AGNs, e.g. \citealt{oleary2006,miller2009,mckernan2012,antonini2016,bartos2017,stone2017,mckernan2018,rasskazov2019,arcasedda2020,arcasedda2020b,yang2019,Yang2019b,Tagawa2020a,Tagawa2020b,Perna2021a,Perna2021b,Zhu2021,Tagawa2021a,Tagawa2021b}) are
dominated by binaries formed via dynamical interactions. In addition to providing important clues on the formation channels of the CO binaries, host galaxy identification is key in order to measure the Hubble constant.

\begin{figure}
\includegraphics[width=0.95\columnwidth]{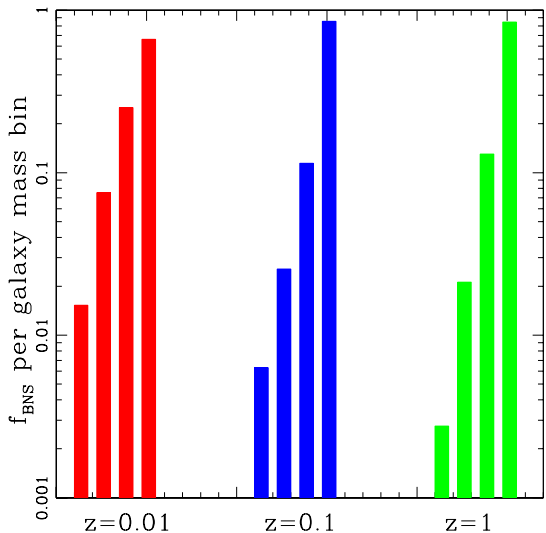}
\vspace{-0.8in}
\caption{The fraction of BNS merger events in
galaxies grouped in 4 mass bins (galaxy {\em stellar} mass). From left to right at each of the 3 redshift snapshots:\
  $8 < \log[{\cal M_{\rm gal,1}}/{\rm M}_\odot ]< 8.75$,
$8.75 < \log[{\cal M_{\rm gal,2}}/{\rm M}_\odot ]< 9.5$,
$9.5 < \log[{\cal M_{\rm gal,3}}/{\rm M}_\odot ]< 10.25$, and
$10.25 < \log[{\cal M_{\rm gal,4}}/{\rm M}_\odot ]< 11$. The relative
fractions are directly proportional to the mass bin, reflecting the fact that BNS mergers are produced without biases with respect to the galaxy mass, per unit mass. Note that the fractions in the various mass bins are normalized to 1 for each redshift snapshot; hence the relative fractions should only be compared within the same snapshot. }
\label{fig:fBNS}
\end{figure}

\begin{figure}
\includegraphics[width=\columnwidth]{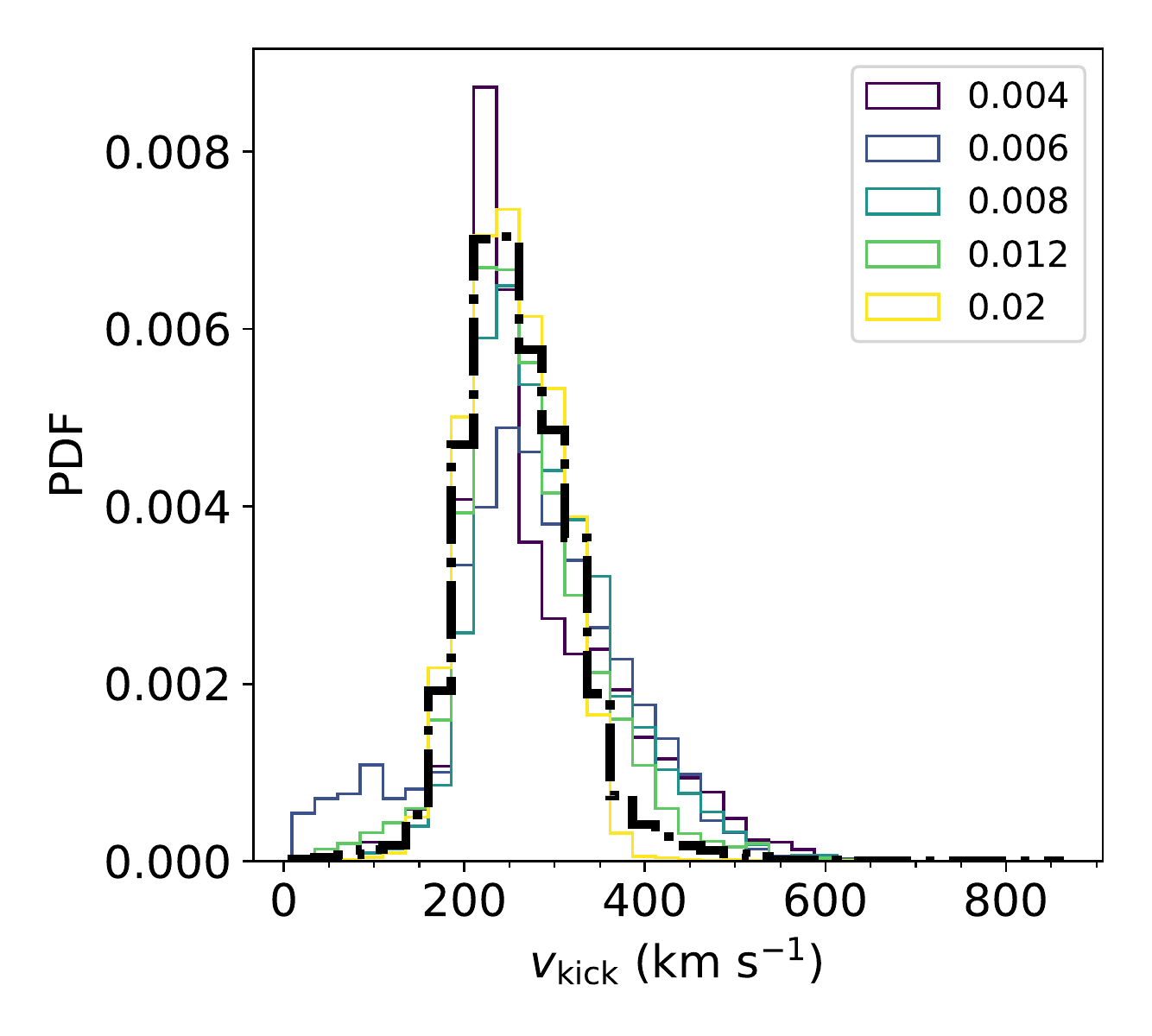}
\caption{
Probability distribution function (PDF) of the binary kick velocities in the centre of mass. Black dash-dotted thick line: kick velocities at $z=0.01$ from the TNG50 simulation. Solid thin lines from yellow to blue: kick velocities at $z=0.01$ divided by the metallicity of the progenitor binary star ($Z=0.02,$ 0.006, 0.008, 0.012 and 0.02).}
\label{fig:vel}
\end{figure}

\begin{figure}
\includegraphics[width=0.95\columnwidth]{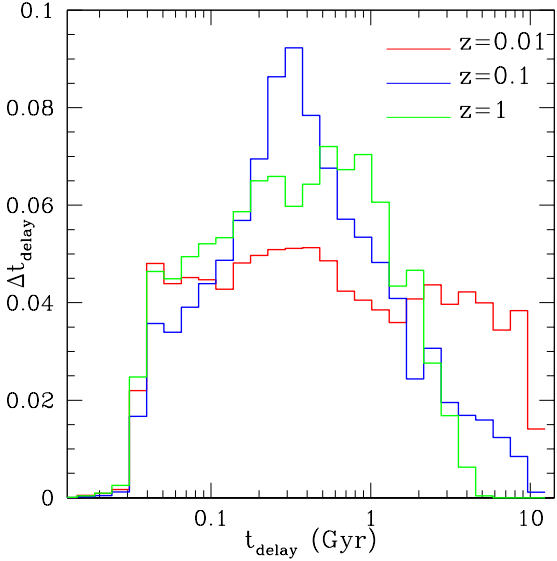}
\vspace{-0.6in}
\caption{Distribution of the binary time delays at the three redshift snapshots of our study. These represent the times elapsed between BNS formation and BNS merger due to GW emission.}
\label{fig:tdel}
\end{figure}

\begin{figure*}
\includegraphics[width=0.95\textwidth]{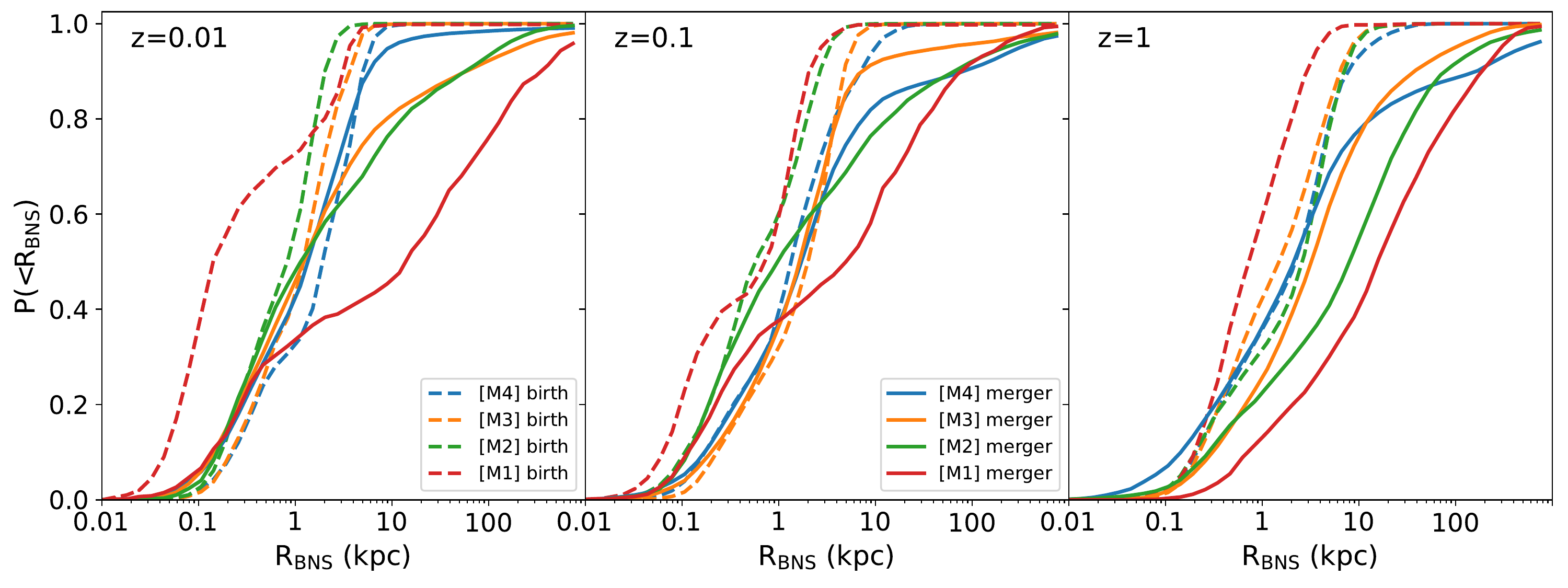}
\caption{Radial distributions of BNS birth sites (dashed lines) and merger sites (solid lines) within their host galaxies. From left to right, the panels display the distributions at three representative redshifts from the Illustris TNG50 simulation. For each redshift, the distributions are displayed for galaxies in 4 {\em stellar} mass bins: [M1]: $8 < \log[{\cal M_{\rm gal,1}}/{\rm M}_\odot ]< 8.75$, 
[M2]: $8.75 < \log[{\cal M_{\rm gal,2}}/{\rm M}_\odot ]< 9.5$,
[M3]: $9.5 < \log[{\cal M_{\rm gal,3}}/{\rm M}_\odot ]< 10.25$ and
[M4]: $10.25 < \log[{\cal M_{\rm gal,4}}/{\rm M}_\odot ]< 11$.   }
\label{fig:radii}
\end{figure*}

Observational and theoretical studies of the host galaxies of binary compact objects pre-date the detection of gravitational waves (e.g. \citealt{Perna2002,voss2003,Bloom2003,Fan2005,Belczynski2006gal,Covino2006,Nakar2006,Oshaughnessy2008,Piranomonte2008,Guetta2009,Davanzo2009,Antonelli2009,Berger2010,Kopac2012,Margutti2012,fong2013,Fong2015}). Since the short gamma-ray bursts (SGRBs) have
long been suspected to be associated with BNS and neutron star--black hole (NSBH) mergers, studies of the galaxy hosts have provided
remarkable clues onto the evolutionary channels leading to the observed SGRBs. Events occurring in elliptical galaxies are more likely associated 
with binaries merging after a long delay time, whereas associations with star forming regions in spiral and starburst galaxies indicates a more prompt merger.

Since the detection of gravitational waves (GWs) from merging compact objects, there has been  a revival of host galaxy studies to further understand the astrophysical origin of 
the detected events, especially when an electromagnetic counterpart accompanies the GW detection.  Most recent works have moved beyond the use of 
semi-analytical models of galaxy population, and rather used galaxy catalogues generated from large-scale numerical simulations, which have been coupled with the results of population synthesis calculations of binary evolution.  This has allowed to explore in better detail the dependence
of binary mergers of different types of compact objects on properties of their host galaxies such as the stellar mass, star formation rate, metallicity, and colors \citep{Cao2018,Adhikari2020,Chu2021,Rose2021}. 
\citet{Mapelli2018}, drawing upon  galaxies from the cosmological box Illustris-1 \citep{vogelsberger2013}, found that BNS mergers tend to form 
and merge in galaxies with stellar mass $\sim 10^9-10^{12}$~M$_\odot$, while NSBH and binary black holes preferentially form in lower mass galaxies $(<10^{10}$~M$_\odot)$
but a fraction of them merge in more massive galaxies due to the longer time delays.
\citet{Artale2019}, using galaxy renderings from the {\textsc{eagle}} simulation \citep{Schaye2015} , found that, for all types of CO mergers and for a wide range of
redshifts between 0 and 6,  there is a strong correlation between the binary merger rate and the stellar mass of the host galaxy.  They further found that
at low redshift early-type galaxies give a larger contribution to the merger rate density than late-type galaxies, while the trend reverses at redshifts higher than
about 1.

In order to be able to compare the results of simulations with actual data, and hence verify/dispute/discriminate among various CO models, it is of paramount
importance to understand whether the population of electromagnetically-detected CO mergers, which is the one allowing host galaxy
identification, is biased with respect to the underlying merging population.  The longer wavelength afterglow radiation following the prompt $\gamma$-rays,
which is key to galaxy  association, depends on the density of gas in the medium. Hence galaxies of different masses, and different merger locations within
the same galaxy, will produce different brightness for the afterglow radiation, even for similar intrinsic properties of the  source (that is jet energetics,
microphysical parameters associated with the afterglow emission,  and viewing angle with respect to the observer).  Whether this may result in a bias
of the galaxy-detected BNS population (and, if so, by what extent), is an open question which we address in this work.

More specifically, here we couple the computation of stellar binaries leading to BNS mergers performed with the code {\sc mobse} \citep{giacobbo2018}, with the galaxy
catalogue from the Illustris TNG50 simulation \citep{Pillepich2019}, and state-of-the-art numerical modeling of the afterglow radiation from BNS mergers \citep{Lazzati2018},
to predict the broadband emission properties of the BNS merging population, for three representative redshifts ($z=0.01,0.1,1$), and for a range of galaxy masses.
We predict the fraction of BNS mergers expected to yield detectable radiation in various representative bands, and we investigate the extent to which
the observable population is biased with respect to the intrinsic sample. Our paper is organized as follows: in Sec.~2 we summarize the various ingredients
of the modeling; Sec.3 describes the results of the simulated population of electromagnetic (EM) sources, and analyzes biases with respect to the intrinsic one.
We summarize in Sec.~4.

\begin{figure*}
\includegraphics[width=0.95\textwidth]{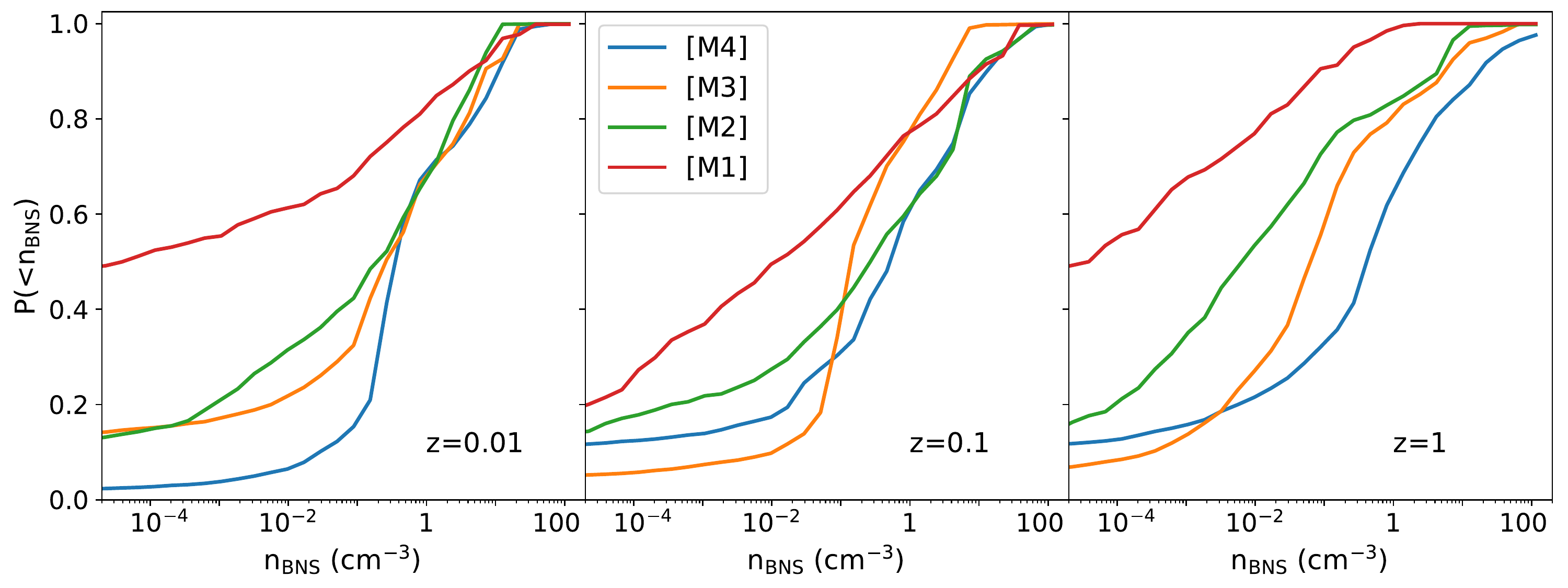}
\caption{Distribution of ISM densities 
at the BNS merger site locations. From left to right, the panels display the distributions at three representative redshifts from the Illustris TNG50 simulation. For each redshift, the distributions are displayed for galaxies in 4 {\em stellar} mass bins: [M1]: $8 < \log[{\cal M_{\rm gal,1}}/{\rm M}_\odot ]< 8.75$, 
[M2]: $8.75 < \log[{\cal M_{\rm gal,2}}/{\rm M}_\odot ]< 9.5$,
[M3]: $9.5 < \log[{\cal M_{\rm gal,3}}/{\rm M}_\odot ]< 10.25$ and
[M4]: $10.25 < \log[{\cal M_{\rm gal,4}}/{\rm M}_\odot ]< 11$.  }
\label{fig:densities}
\end{figure*}

\section{Methods}

\subsection{Population synthesis calculations of binary stars leading to BNS mergers}

As discussed in the introduction, binary compact objects, and in particular BNSs of interest here, 
can form via a variety of formation channels, which are favoured in different environments. 
In this work, we specifically focus on BNS systems formed from isolated binary star evolution  via common envelope.

The evolution of stars in binary systems, and hence the properties of BNSs at the time of mergers, have been the subject of
numerous investigations using population synthesis codes (e.g. \citealt{tutukov1973,portegieszwart1998,belczynski2002,voss2003,belczynski2007,podsiadlowski2004,oshaughnessy2010,eldridge2016,marchant2016,belczynski2016,demink2016,Belczynski2017,stevenson2017,Mapelli2018,giacobbo2018,vignagomez2018,neijssel2019,tang2019,tanikawa2020,mandel2021}), coupled with either semi-analytical prescriptions for galaxy
modeling and evolution (e.g. \citealt{Perna2002,Belczynski2006gal}), or with results from numerical simulations of galaxy formation (e.g. \citealt{Artale2019,Artale2020,Artale2020b, Briel2021,Chu2021,Mandhai2021}).  The predicted global
rates suffer from rather large uncertainties due to a combination of not-well constrained model inputs (i.e. the  initial mass function, the metallicity
and its cosmic evolution, the star formation rate, the natal kick prescription), and the physics of common envelope evolution \citep[e.g.,][]{dominik2013,Mapelli2018,chruslinska2018,chruslinska2019,Santoliquido2021}. 
In this work,  to minimize the influence  on our results from the uncertain, redshift-dependent input parameters, we focus on selected representative
redshifts slices, and on the statistical properties of the galaxies at those redshifts as derived from state-of-the-art cosmological simulations (Sec 2.3).

Isolated binary evolution is modeled using the population synthesis code {\sc mobse} \citep{giacobbo2018}.  In the following we summarize
the key elements and assumptions of this code, while more details can be found in the papers cited above. 
The mass loss rate of a massive hot star of metalllicity $Z$  is modeled according to the prescription $\dot{M}\propto Z^\beta$, where $\beta$ is a parameter
dependent on the Eddington ratio \citep{giacobbo2018}. 
The initial mass function follows the Kroupa law \citep{kroupa2001} in the high mass range,  that is $dN/dM\propto M^{-2.3}$. 
The orbital periods, eccentricities and mass ratios of the massive binary stars are drawn from \cite{sana2012}, yielding the distribution
$\mathcal{F}(q) \propto q^{-0.1}$ with $q\in [0.1-1]$ for the
mass ratio $q=m_2/m_1$; the orbital period $P$  is drawn from $\mathcal{F}(\Pi) \propto \Pi^{-0\
.55}$ with $\Pi = \log{(P/\text{day})} \in [0.15 - 5.5]$ and the eccentricity $e$ from $\mathcal{F}(e) \propto e^{-0.42}~~\text{with}~~ 0\le\
 e \leq 0.9$.


The metallicity distribution comes directly from the Illustris TNG50 simulation.  
Our simulation grid with {\sc mobse} has 12 values of metallicity between $Z=0.0002$ and $Z=0.02$.  
Mass transfer via Roche lobe overflow is modeled according to the prescription of \citet{hurley2002}, yielding a nearly
conservative mass transfer if the accretor is a non-degenerate star. 

The assumed functional form of the neutron star (NS) kick distribution is 
\begin{equation}
\label{eq:kicks}
    v_{\text{kick}} = f_{\text{H05}}\frac{m_{\text{ej}}}{\langle m_{\text{ej}}\rangle}\frac{\langle m_{\text{NS}}\rangle}{m_{\text{rem}}},
\end{equation}
where 
$m_{\rm ej}$ is the mass of the ejecta, $m_{\rm rem}$ is the mass of the compact remnant, $\langle m_{\text{NS}}\rangle$ is the average NS mass and $\langle m_{\text{ej}} \rangle$ is the average mass of the ejecta associated with the formation
of a NS of  mass $\langle m_{\text{NS}}\rangle$ from single stellar evolution. Finally, $f_{\text{H05}}$ is a random value extracted from a Maxwellian distribution with one-dimensional root mean square $\sigma_{\rm 1D} = 265$~km~s$^{-1}$. We use a Maxwellian  distribution with  $\sigma_{\rm 1D} = 265$~km~s$^{-1}$ because this matches the proper motions of young pulsars in the Milky Way \citep{hobbs2005}. We refer to \cite{giacobbo2020} for more detail on the kick model. 

\subsection{Mergers in host galaxies from the Illustris TNG50 simulation}

The host galaxies of binary compact objects 
are selected from the cosmological
magnetohydrodynamical simulation TNG50 \citep{Pillepich2019}, the third and final of the IllustrisTNG project.
TNG50 initially contains $2160^3$ dark matter particles and the same number in gas cells
 in a volume of  (50~Mpc)$^3$.
It has an average spatial resolution of $\sim 70-140$~pc, and is able to resolve baryonic masses down to
$8.5\times 10^4$~M$_\odot$. The simulation includes several key elements of sub-grid physics, such as star formation, cooling, supernova and AGN feedback, accretion and mergers, and formation of supermassive black holes \citep{vogelsberger2013,pillepich2017}.
At a redshift of 1, Illustris TNG50 samples about 6500 galaxies with stellar mass larger than
$10^8$~M$_\odot$ at an unparalleled level of detail, resolving internal structure of galaxies and providing 
insight into their chemo-dynamical evolution. The simulation spans a large swath of cosmic history, from very high redshifts
 to the present time. 
 The initial conditions of the simulation series have been created at $z=127$.  In this work, we focus on three redshift snapshots: $z=0.01, 0.1, 1$. 
These encompass the range in which (at least the brightest) electromagnetic counterparts from short GRBs can be detected,
while the lowest redshift, corresponding to a distance of $\sim 45$~Mpc, is within the current LIGO--Virgo horizon. 
Analysis of the simulated EM counterparts at this redshift will allow us to evaluate the probability of detecting an EM counterpart to a GW-detected BNS merger.

The catalogues of merging BNSs are coupled with the galaxies in the Illustris TNG50 simulation
following the formalism developed by \citet{mapelli2017}. The {\sc mobse} simulations provide, for each
binary system which merges in less than the Hubble time, the NS masses and the delay time $t_{\rm delay}$ between the formation of the progenitor
stars and the BNS merger. For a given mass particle $M_{\rm TNG50}$ formed at a redshift $z_{\rm TNG50}$ and with metallicity $Z_{\rm TNG50}$ in the Illustris TNG50 simulation, we associate a number $n_{\rm BNS}$ of BNSs from progenitors stars with the closest metallicity  to the {\sc mobse} tables ($Z_{\rm TNG50}\sim Z_{\rm BSE}$) as
\begin{equation}
n_{\rm BNS} = N_{\rm BSE}\frac{M_{\rm TNG50}}{M_{\rm BSE}} f_{\rm corr} f_{\rm bin}
    \label{nbns}\,.
\end{equation}
Here, $N_{\rm BSE}$ is the number of merging BNS within the simulated sub-set of initial stellar mass $M_{\rm BSE}$ with {\sc mobse}, $f_{\rm bin}=0.4$ is the assumed fraction of stellar mass in binaries, and $f_{\rm corr}=0.285$ corrects for the fact that only binaries with primary mass larger than $5$~M$_\odot$ are simulated.
The lookback time of the merging BNSs in the Monte Carlo selected sample, $t_{\rm merg}$, is given by $t_{\rm merg}=t_{\rm form}-t_{\rm delay}$, where $t_{\rm form}$ is the looback time at which the particle in the Illustris TNG50 simulation has been formed.

\begin{figure}
\includegraphics[width=0.95\columnwidth]{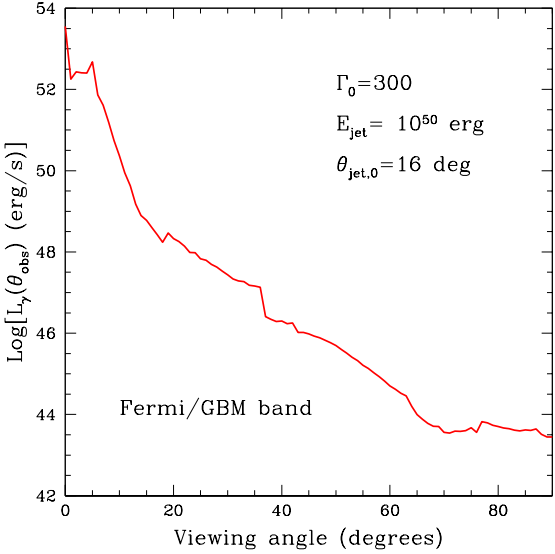}
\vspace{-0.8in}
\caption{Isotropic peak luminosity in the {\em Fermi/GBM} band as a function of the viewing angle with respect to the jet axis. The initial jet parameters are those typical of a short GRB, and the luminosity at $\theta_{\rm obs}\sim 25-30$~deg matches that of GRB170817A.}
\label{fig:prompt}
\end{figure}

For each of the three representative redshift snapshots from the TNG50 simulations, we divide a sample of  randomly selected galaxies into four {\em stellar} mass bins:
$8 < \log[{\cal M_{\rm gal,1}}/{\rm M}_\odot ]< 8.75$ (referred to M1 in the figures label), 
$8.75 < \log[{\cal M_{\rm gal,2}}/{\rm M}_\odot ]< 9.5$ (referred to as M2),
$9.5 < \log[{\cal M_{\rm gal,3}}/{\rm M}_\odot ]< 10.25$ (M3) and
$10.25 < \log[{\cal M_{\rm gal,4}}/{\rm M}_\odot ]< 11$ (M4).
This sub-division allows us to investigate whether  
there is any mass dependence in the statistical properties of the BNS mergers, and, if so, whether
this may affect the BNS observability in EM emission. The relative fraction of BNS merger events in each mass galaxy bin is displayed in Fig.~\ref{fig:fBNS}
for the sample of galaxies at each redshift snapshot of our study.  {In each considered redshift bin, the relative fraction of BNS mergers scales with the mass of the host galaxy: the most massive galaxies tend to host more BNS mergers with respect to low mass galaxies. As already discussed in previous work \citep{Mapelli2018b,Artale2019,Artale2020}, the formation rate of BNSs is mainly sensitive to the star formation rate of the host galaxy and barely affected by metallicity. Moreover, the delay time of a coeval BNS population scales as $\sim{t^{-1}}$ \citep{dominik2012,Mapelli2018b,Mapelli2019}. Hence, the most massive galaxies in our sample, which, at high redshift, are also associated with the highest star formation rate, host the largest number of mergers \citep{Artale2020}.}

{At low redshift ($z=0.01$), the star formation rate of the most massive galaxies is significantly quenched \citep[e.g.,][]{Moffett2016}. Hence, the relative fraction of BNS mergers   in low-mass galaxies becomes more important at low redshift. However the amount of mass locked in the most massive galaxies at low redshift is so large \citep{Moffett2016} that massive galaxies still dominate the merger rate per galaxy \citep[e.g., Figure 5 of][]{Artale2020}.}

Fig.~\ref{fig:vel} shows the velocity distribution of the center of mass of the binaries ($v_{\rm kick}$) at  redshift $z=0.01$.  
{To make this Figure, we considered the centre-of-mass velocity of each binary system after the second supernova explosion, as described in \cite{hurley2002}. The centre-of-mass velocity $v_{\rm kick}$ is very sensitive to the natal kick model we adopted in our simulations \citep{giacobbo2020}.  Systemic velocities at  $z=0.01$ peak at $\sim{220}$ km s$^{-1}$ and are mainly the fingerprint of BNSs produced by progenitors with solar and slightly sub-solar metallicity (see \citealt{giacobbo2020} for more details).} 

Fig.~\ref{fig:tdel} shows the time delay distributions at the three snapshots. It is evident that the time delay is generally longer for binaries merging at lower redshifts. 
The delay time of a coeval population of BNSs is approximately $dN/dt \propto t^{-1}$.  
However what is observed at low redshift is not a single coeval population of BNSs, but rather a piling up of different populations with different ages.
Most of the BNSs merging at $z\sim 0.01$ come from the tail of the $t^{-1}$ distribution for large $t$ and are the result of multiple episodes of star formation across the cosmic time {(see Figure 4 of \citealt{Mapelli2018b})}. 
On the other hand, most of the BNSs merging at $z\sim 1$ (cosmic time closer to the peak of the star formation rate) formed in the most recent burst of star formation and come from the low $t$ end of the $t^{-1}$ distribution. 

The velocity and time delay distributions are
used to compute the distribution of the BNS merger sites, with their 
 birth sites computed as described below. 
 For each binary characterized by a center of mass velocity $v_{\rm kick}$ and a merger time $t_{\rm delay}$, its merger location is  computed by numerically integrating its orbit in the potential (stars + gas + dark matter) of the corresponding host galaxy.
The direction of the velocity is assumed to be randomly distributed with respect to its position within the galaxy, and each orbit is integrated for a time corresponding to the delay time between the formation of the binary and its merger.  We note that this procedure implicitly assumes that the galaxy potential does not evolve during the travel time of the BNS between formation and merger. Therefore, we do not account for the changing potential that would occur if the host galaxy were to merge with another galaxy during the BNS travel time. 
Various studies show that the number of galaxy mergers as a function of the cosmic time depends on both galaxy mass and redshift.  Massive galaxies are more likely to be affected by major mergers during their formation assembly at high redshifts, decreasing their probability at lower redshifts \citep[see e.g.,][]{Genel2010,RodriguezGomez2015}. In particular, those numerical studies with the Illustris simulation find that the rate of a major merger (i.e. between two massive galaxies with mass ratio $<1/4$) is $< 0.02$~Gyr$^{-1}$ at redshifts $z<1$.

Fig.~\ref{fig:radii} shows the radial distribution of BNS birth sites (dashed lines) and merger sites (solid lines) for the galaxies in the four mass bins described above, and for  the three redshift snapshots in our study. The distribution of the merger sites depends on the distribution of birth sites, the velocity distribution of the binaries, their travel time up to merger (time delay), and the potential of the galaxies in which they are hosted. For a given merger location, kick velocity and time delay, the BNSs will travel more 
in the smaller  galaxies, due to their smaller gravitational potentials. This trend is evident in Fig.\ref{fig:radii}: for example, at all redshift snapshots, about 20\% of BNSs from
the smaller galaxy group (M1 in the figure) is found at distances $\gtrsim 100$~kpc. On the other hand, a smaller fraction reaches larger radial distances in the most massive galaxy groups. These results are in broad agreement with those of \citet{Belczynski2006gal}. 
We should note that these radial distributions  (and thus offset distributions) are not meant to be for direct comparison for observations of short GRBs. This is because merging BNSs can only be localized if their emission as short GRBs can be detected. However, detections are biased towards the brightest events, which are on average the ones in the innermost parts of the galaxies (see also discussion in \citealt{Mandhai2021} on this point).

The distributions of the interstellar medium (ISM) densities corresponding to the merger sites distributions of Fig.~\ref{fig:radii} are displayed in Fig.~\ref{fig:densities}. 
Each value of the distribution is essentially determined by the mass of the galaxy (less massive galaxies have on average lower ISM densities), and by the location of the BNS merger within that galaxy. Both effects contribute to an ISM density distribution which is considerably biased towards low values in the small galaxy sample, compared to the larger galaxies. As discussed in the following, the density distributions hence bias the observability of afterglow emission from BNSs mergers towards larger galaxies.

\begin{figure*}
\includegraphics[width=0.95\textwidth]{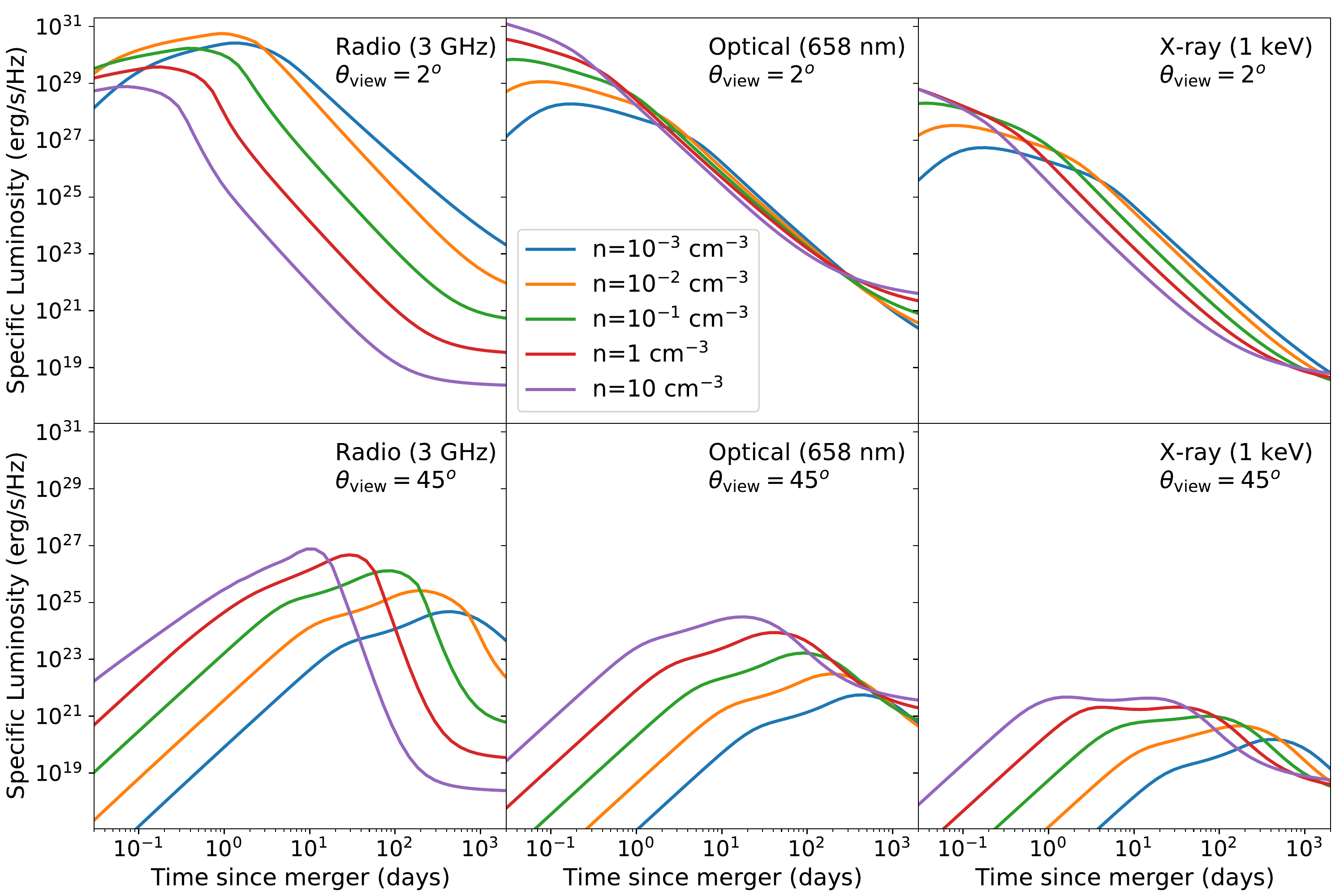}
\caption{ISM density dependence of the afterglow luminosity density in three representative bands (radio, optical, X-rays, from left to right) and for an
on-axis viewing angle ($\theta_{\rm view}=2$~deg, top panels), or for a more generic and likely viewing angle,
 $\theta_{\rm view}=45$~deg (bottom panels) with respect to the jet axis. Note that the scale on the $y-$axis is the same in all the panels, highlighting the significant drop in luminosity at larger viewing angles.}
\label{fig:luminosities}
\end{figure*}

\subsection{Electromagnetic radiation from BNS mergers}

The association between GW170817 and GRB170817A has confirmed that BNS
mergers yield transients with properties consistent with those of the cosmological GRBs
\cite{Lazzati2018,Mooley2018,Salafia2019}. Therefore, we model the electromagnetic (EM) counterparts to the BNS mergers
following the theory developed for SGRBs, and refined for the modeling of GRB170817A.

More specifically, we consider as input data a
`structured' jet, that is a jet whose angular properties (energy distribution and Lorentz factor) have been molded by the interaction with the ejecta of a BNS merger. The jet simulations, which were performed by \citet{Lazzati2017} with the code {\sc flash} \citep{Fryxell2000}, start with a top-hat jet of angular size {$16$~deg}, as typically inferred for short GRBs \citep{Fong2015}, energy $L_j=10^{50}$~erg~s$^{-1}$, engine duration 1~sec, and Lorentz factor $\Gamma_0=300$, and evolve the jet within an ejecta described by 
a density profile $n=n_0 (r/r_0)^{-2}e^{-r/r_0}$, with $r_0=10^{18}$~cm and $n_0=10^6$~cm$^{-3}$, made to reproduce results from realistic, GRMHD simulations (e.g. \citealt{Kawamura2016,Ciolfi2017,Camelio2021,Ruiz2020,Murguia2021,Foucart2019,Kiuchi2018,Most2019,Radice2017}).

The prompt radiation is computed assuming that the outflow dissipates its internal energy at some radius $R_{\rm rad}$ from the engine, and the observed bolometric flux is calculated by adding the contributions of the local emission from the whole emitting surface, boosted to the fourth power of the Doppler factor $[1-\beta \cos\theta]^{-1}$, where $\beta$ is the speed of the jet divided by the speed of light, and $\theta$ is the angle that the local photon makes with the normal to the emitting surface.
The spectrum is assumed to be a Band one \citep{Band1993}, that is a broken power law with photon spectral indices of $\alpha_{\rm ph}=-1$ below 
a peak of 500~keV in the comoving frame, and
 $\beta_{\rm ph}=-2.5$ above it. The results of the light curve computation are displayed in 
 Fig.~\ref{fig:prompt}, which specifically shows the peak luminosity in the {\em Fermi/GBM} observation band. At a viewing angle of about 25-30~deg, the luminosity matches that observed 
in GRB170817A.

\begin{figure}
\includegraphics[width=0.95\columnwidth]{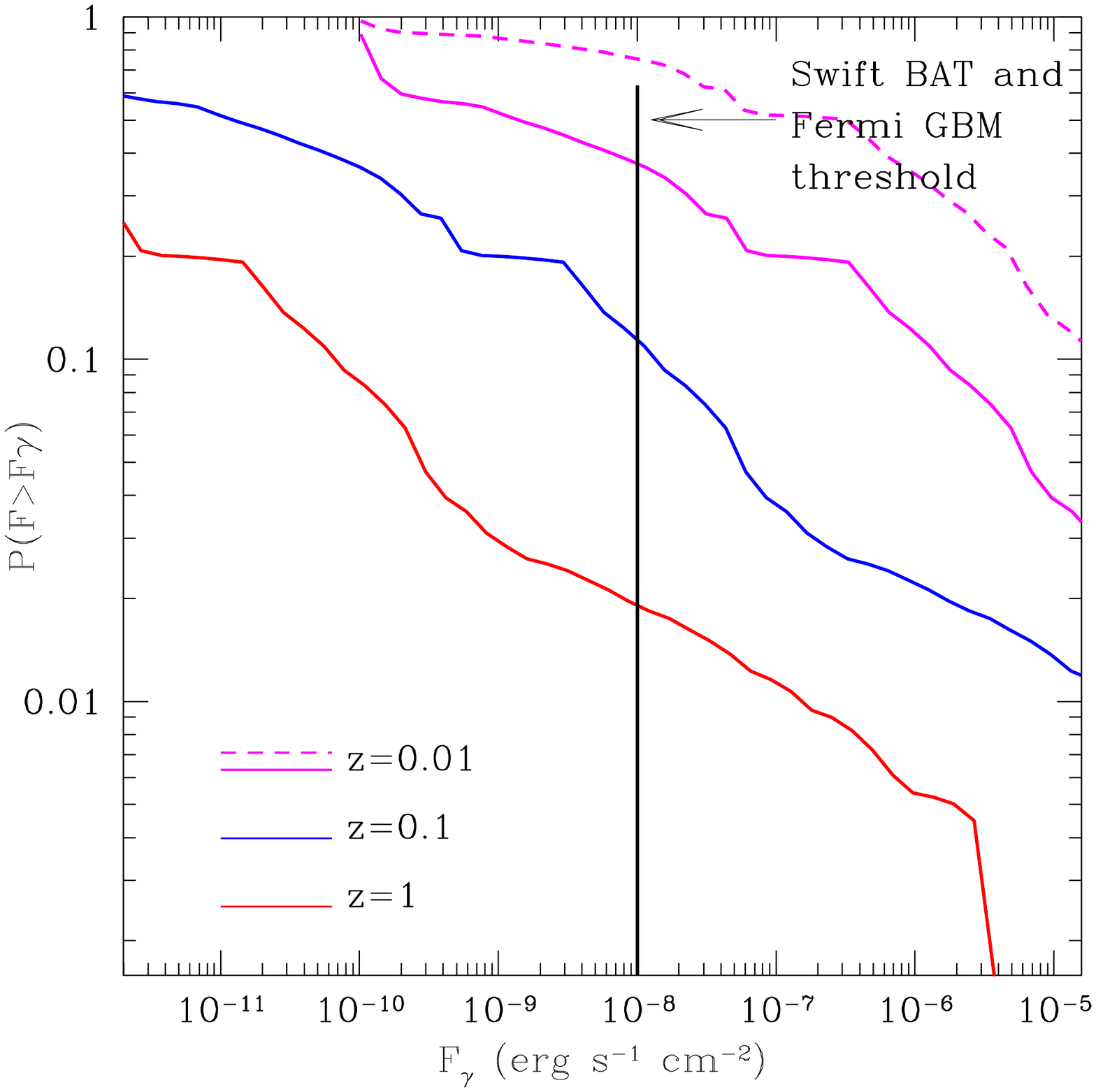}
\vspace{-0.8in}
\caption{The fraction of BNS merger events, with jet properties similar to those of GRB170817A,
which have peak flux in $\gamma$-rays larger than $F_{\gamma}$, as a function of
$F_{\gamma}$ and for four different distances. The smallest distance of $z=0.01\sim 45$~Mpc is  within the current horizon of LIGO--Virgo for BNS mergers.
The orientation of the jet with respect to the observer is assumed to be random on the sky
except for $z=0.01$, where both the random orientation (solid line) and the GW-detected case (dashed line) have been simulated.
As expected, at cosmological distances the detectable 
fraction of $\gamma$-ray events with current detectors is rather small,
since the luminosity drops rapidly with angle (cfr. Fig.~\ref{fig:prompt}) and hence only jets observed at small enough viewing angles are
luminous enough to allow detection. Note that the displayed probabilities do not account for the field of view of the instruments (9.5~sr for {\em Fermi} and 1.5~sr for {\em Swift}.}
\label{fig:fluxes-gamma}
\end{figure}

After the structured jet has released the prompt emission, it propagates into the external medium where it generates an external shock and eventually dissipates its energy. In the process, particles accelerated by the shock emit synchrotron radiation thanks to the magnetic fields generated downstream of the shock. This radiation, which spans the wide electromagnetic range from the X-rays to the radio, is the so-called afterglow. 
We compute afterglow light curves and spectra using standard techniques \cite{Rossi2004,Sari1998,Granot2002,Panaitescu2000}, involving the integration of the local emission over the equal arrival times, for an observer located at  any line of sight with respect to the jet axis. The afterglow radiation depends on the microphysical shock parameters that describe the particle distribution and magnetic field intensity downstream the shock. These are parameterized via the fraction $\epsilon_e$ of energy which goes to the electrons, and the fraction $\epsilon_B$ of energy which goes into the magnetic field. Here we adopt the values of these parameters which provided the best fit to the broadband light curves of GRB170817A:  $\epsilon_e=0.03$ and $\epsilon_B=0.003$, and the index of the electron distribution 
$p=2.13$.
With these fixed, we ran a grid of light curves for a range of number densities of the interstellar medium between $10^{-6}$~cm$^{-3}$ and $100$~cm$^2$.

Fig. \ref{fig:luminosities} shows, for two representative viewing angles $\theta_{\rm view}=2$~deg (top panels) and $\theta_{\rm view}=45$~deg (bottom panels)
with respect to the jet axis, the afterglow light curves in three representative bands (X-rays, optical and radio) for a range of number densities in the interval $n=[10^{-3}-10]$~cm$^{-3}$, which encompasses a good fraction of the density values expected at the BNS merger sites (cfr. Fig.~\ref{fig:densities}).  A visual comparison between 
{upper and lower} panels (note the same scale on the $y-$axis) immediately highlights 
the strong dependence of the luminosity on the viewing angle with respect to the line of sight to the observer. Hence, GW-detected BNS mergers, which are more likely to be seen at larger viewing angles, are expected to have on average significantly dimmer luminosities with respect to the cosmological SGRBs, which can only be detected when close to on-axis.

An important feature to notice of the off-axis afterglow light curves is that their peak brightness is much later than for the on-axis GRBs. This is because the maximum luminosity is achieved when the Doppler factor of the emitting jet becomes on the order of $\Gamma\sim 1/\theta_{\rm view}$, and hence radiation from the more energetic central regions of the jet can reach the observer. The specific time at which this happens depends on the medium density. At higher densities, the blastwave decelerates more quickly, and emission from the central jet regions enter the line of sight at earlier times. At 45 degrees angles, all frequencies shown are above the self-absorption frequency, and therefore the peak luminosity is larger for larger densities in all the bands, reflecting the larger fraction of emitting electrons for higher densities (see e.g. \citealt{Sari1998}). On the other hand, the radio band is below the self-absorption frequency for the on-axis light curves, causing the radio emission in denser media to be dimmer due to the larger opacity of the blast wave.

Another feature of the light curves which is worth noticing (and which will play a role in the interpretation of the statistical results for the EM emission), is the fact that, while at small viewing angles (upper panels) the peak specific flux of the radio emission is comparable to that in the optical, and only moderately larger than the peak flux in X-rays, at large viewing angles (bottom panels), the radio brightness is significantly larger than the optical, and even more so than the X-ray one. This is because the physical reason for the peak of the light curves in different bands is different for on-axis and off-axis observers. For on-axis observers the peak is due to the fact that the spectrum moves towards lower frequencies as the fireball decelerates, maintaining a constant peak specific flux \citep{Sari1998}. Higher frequencies peak earlier in time but, as long as there is no self-absorption, all bands peak with the same specific flux. For off-axis viewers, instead, the peak is due to the entering of the core emission within the line of sight. While the wings contribute to the observed emission, the core dominates at all bands after it becomes visible. In this case the peak happens almost simultaneously in all bands, since it is a geometric effect. In most relevant cases, since the emission peak is seen at large angles form the core and at late times, the peak frequency is at low frequencies and lower frequencies appear brighter for off-axis observers.
This results in a ratio between the radio and the X-ray peak fluxes which is much larger at larger viewing angles than near the core.

The grid of light curves will be used, together with the information on the BNS sites provided by the calculations described in Sec.~2.1 and 2.2, to predict the statistical properties of the BNS EM counterparts in Sec.3.

\begin{figure*}
	\includegraphics[width=1.0\textwidth]{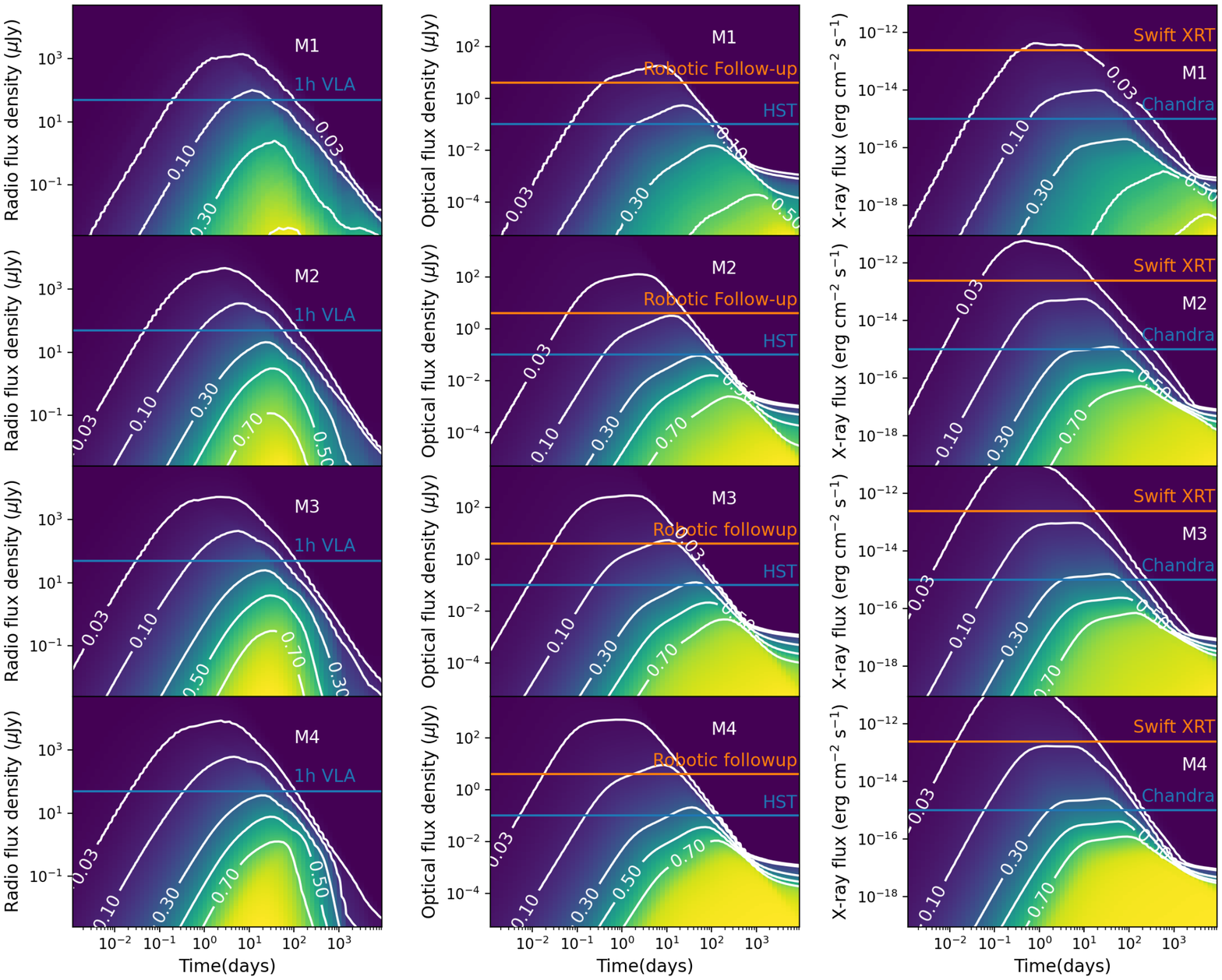}  
	 \caption{Probability of detecting an EM counterpart in three wavelength bands (radio, optical, X-rays from left to right) as a function of time from the merger event, for BNS mergers in galaxies at the redshift snapshot $z=0.01$. From top to bottom, the various panels refer to the four galaxy mass bins of our study. The numbers in the contour lines represent the fraction of  BNS mergers with flux larger than the corresponding value (on the $y$ axis) at the corresponding time to the $x$ axis. Here the viewing angle with respect to the jet axis is drawn from a random distribution.}
    \label{fig:aft_ran}
\end{figure*}

\begin{figure*}
	\includegraphics[width=1.0\textwidth]{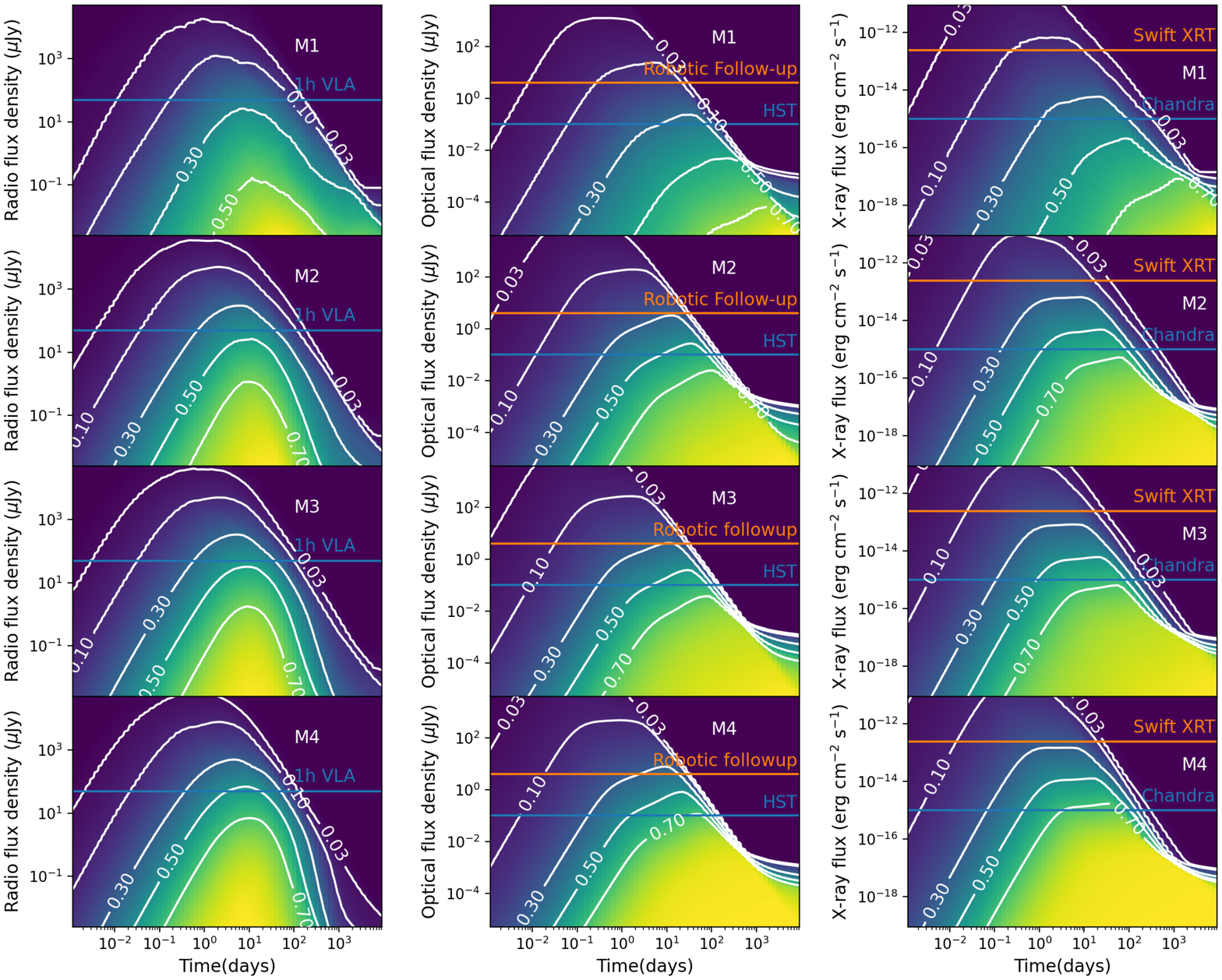}  
	 \caption{Same as Fig.~\ref{fig:aft_ran} but with the jet angle assumed to coincide with the perpendicular to the orbital plane of the binary, and hence the probability distribution for the viewing angle (assuming a GW trigger) given by Eq.~\ref{eq:pobs}.}
    \label{fig:aft_obs}
\end{figure*}

\section{Results: Monte Carlo simulation of the expected electromagnetic source population}

We generate the observable population, at each redshift snapshot and for each
of the four galaxy mass bins,
 by performing  Monte Carlo random realizations
for each of these sub-populations.
For each case we run $10^5$ different random realizations.
The  merger sites are drawn
from  the probability distributions of Fig.~\ref{fig:radii}, and the corresponding ambient densities
from the curves of  Fig.~\ref{fig:densities}. 
The inclination angle of the jet is assumed to be isotropically distributed on the sky.

Fig.~\ref{fig:fluxes-gamma}  shows the distributions of $\gamma$-ray fluxes (prompt emission) from BNS mergers in host galaxies at our three representative redshifts $z=0.01,0.1,1$. 
Since the prompt emission is independent of the ambient medium of the host galaxies, the BNS population
has not been sub-divided by groups in galaxy mass. 
 The relative number of prompt ($\gamma$-ray) events for each galaxy mass bin is simply proportional to the relative number of BNS mergers in that  mass range (cfr. Fig.~\ref{fig:fBNS}). 
 For an immediate gauging of the fraction of detectable events by current $\gamma$-ray detectors, the vertical line indicates the flux value $10^{-8}$~erg~cm$^{-2}$~s$^{-1}$, corresponding to the detection threshold of the {\em Swift} BAT detector for a typical GRB spectrum\footnote{\url{https://swift.gsfc.nasa.gov/about_swift/bat_desc.html}}. In the case of the {\em Fermi} BAT detector, the sensitivity is provided in photon counts\footnote{\url{https://gammaray.msfc.nasa.gov/gbm/instrument/}}, and the conversion to fluence requires a spectral assumption. However, noting that the weakest detected burst in the Fermi catalogue \citep{Bhat2016} has a flux  $2\times 10^{-8}$~erg~cm$^{-2}$~s$^{-1}$, we can assume that the threshold fluxes of the two instruments are roughly comparable.
 The figure shows that the fraction of detectable events is rather small,  $\sim 15\%$, at the highest simulated redshift of $z=1$.  
 The relatively small probability of a prompt $\gamma$-ray detection at this redshift is a result of the fact that, since the direction of the jet axis is assumed to be uncorrelated with the viewing angle, a large majority of events are being caught at large viewing angles, where the luminosity is considerably dimmer (cfr. Fig.~\ref{fig:prompt}).
 
 The smallest redshift of $z=0.01$, corresponding to a luminosity distance of about 45~Mpc, is within the current horizon of LIGO and Virgo to BNS mergers \citep{abbott2018observingscenario}, 
 and it is comparable to the distance of GW170817 \citep{abbottGW170817}. At this distance, 
 about 50\% of the $\gamma$-ray counterparts to BNS mergers are expected to be detected by current satellites. This probability is enhanced if we account for the fact that for BNS mergers whose first trigger is in GWs, the viewing angle may be correlated with the jet axis. 
 More specifically, if the jet axis is in the direction of the orbital angular momentum of the binary, then the distribution of viewing angles will be determined by the detection probability of GW detectors. We hence consider also this situation, which is the relevant one for events within the LIGO-Virgo horizon.
 The detection probability as a function of the angle ${i}$ between observer and rotation axis of the binary can be written as \citep{Schutz2011}
 \begin{equation}
 P({i}) = 0.076076 \;(1+6 \cos^2 i+ \cos^4 i)^{3/2}\,\sin i\,.
     \label{eq:pobs}
 \end{equation}
 For the lowest redshift of $z=0.01$ we ran a Monte Carlo simulation with the assumption that the viewing angle with respect to the jet, that is $\theta_{\rm view}$ is equal to $i$. The results are shown with the dashed line in Fig.~\ref{fig:fluxes-gamma}. 
 It should be noted that, to read off those curves as actual observing probabilities without any prior on the sky localization, they should be corrected for the field of view of the observing instruments. This is 9.5~sr for {\em Fermi} (giving a correction factor of $\sim 0.75$, and 1.5~sr for {\em Swift}
 (giving a correction factor of $\sim 0.1$). 
 
 \begin{figure*}
	\includegraphics[width=1.0\textwidth]{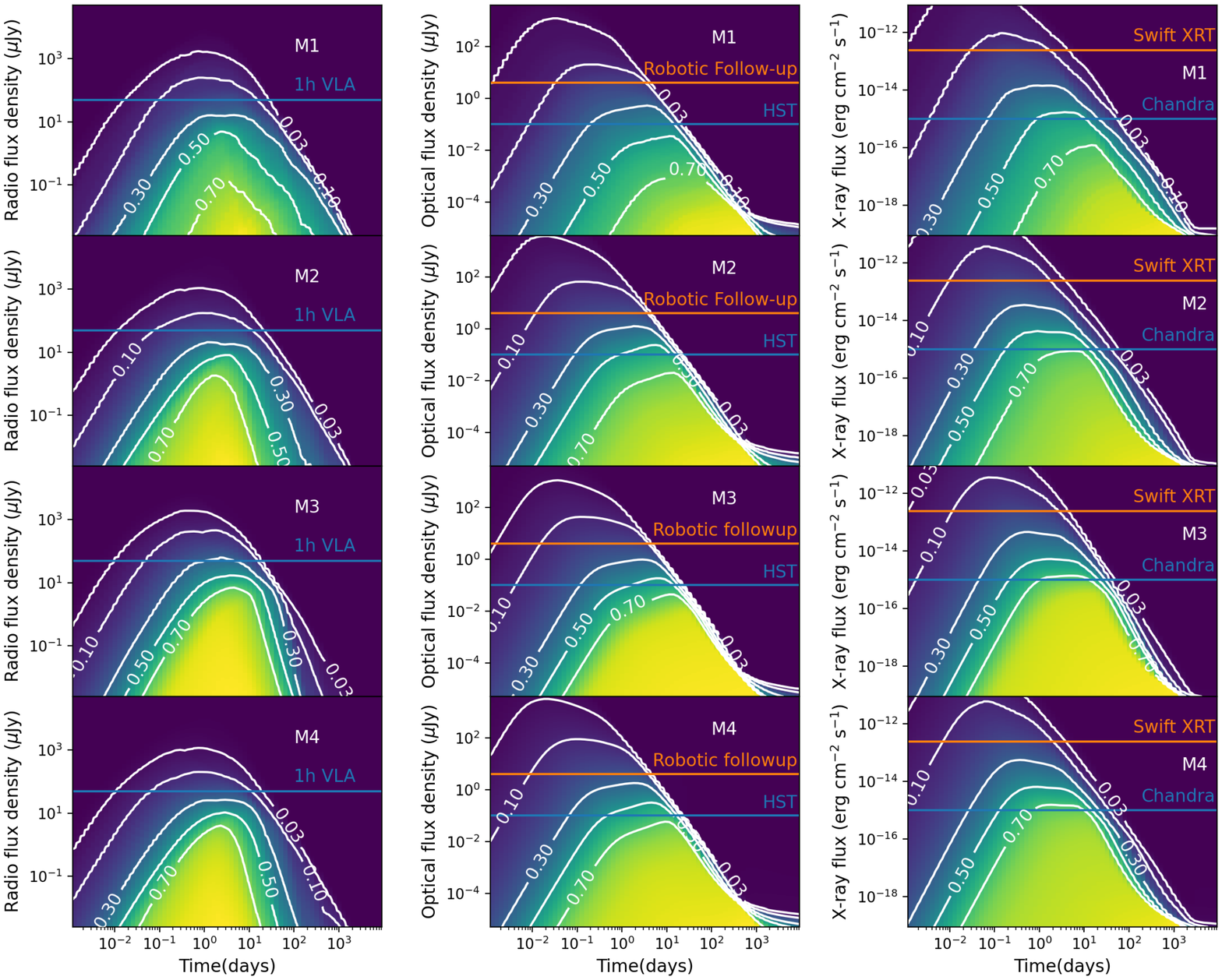}  
	 \caption{Same as Fig.~\ref{fig:aft_ran} but for the sample of galaxies at the redshift snapshot $z=0.1$, and with the condition of a viewing angle (with respect to the jet axis) randomly chosen but such that $\theta_{\rm view}\le \theta_{{\rm max},\gamma}$, where
	  $\theta_{{\rm max},\gamma}$ is the
	   maximum value for which the $\gamma$-ray emission is large enough to trigger  the {\em Swift/BAT} and {\em Fermi/GBM} detectors. At the redshift of $z=0.1$, we find $\theta_{{\rm max},\gamma}\sim 29.4$~deg, for the GRB170817A-like event adopted here.
	   This situation simulates the  one of the standard cosmological SGRBs; that is BNS mergers outside of the current LIGO--Virgo GW horizon, which are routinely triggered  by their prompt $\gamma$-ray emission and later searched at longer wavelengths.}
    \label{fig:aftz0.1}
\end{figure*}
 
 For the longer wavelength (afterglow) radiation, we perform different sets of calculations for the $z=0.01$ snapshot than for the $z=0.1$ and $z=1$ snapshots. For the former, which is well within the current horizon for GW detections, we simulate the two cases discussed above for a GW-triggered event: random viewing angle if the jet direction is uncorrelated with the orbital plane of the binary, and viewing angle drawn from the probability distribution in Eq.~\ref{eq:pobs} if the jet is pointing roughly perpendicular to the orbital plane of the merging NSs.
 On the other hand, for the higher redshift snapshots which are beyond the GW current horizon to BNS mergers, we simulate the current  astrophysical scenario  
  of the 'standard' cosmological short GRBs, which are first triggered in $\gamma$-rays, and then followed at longer wavelengths.
  Hence for these we will restrict the distribution of viewing angles to those for which the prompt emission is above the detection threshold of {\em Swift} and  {\em Fermi}.

\begin{figure*}
	\includegraphics[width=1.0\textwidth]{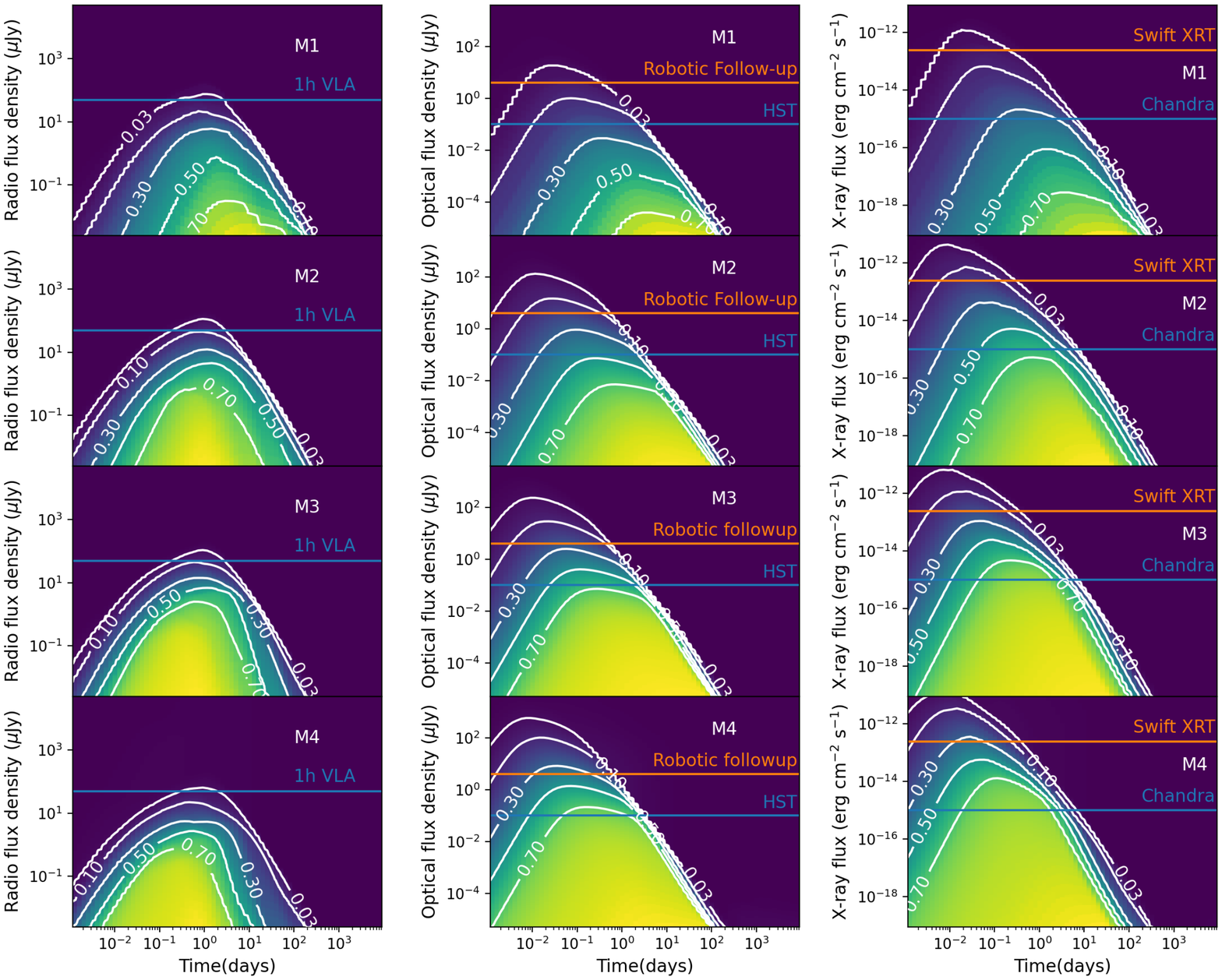}  
	 \caption{Same as Fig.~\ref{fig:aftz0.1} but for the sample of galaxies at the redshift snapshot $z=1$, with the condition of a viewing angle smaller than a maximum value to allow $\gamma$-ray trigger of the event by the
	    the {\em Swift/BAT} and {\em Fermi/GBM} detectors. At the redshift of $z=1$, we find $\theta_{{\rm max},\gamma}\sim 11.5$~deg.}
    \label{fig:aftz1}
\end{figure*}

Figures~\ref{fig:aft_ran} and \ref{fig:aft_obs} show the fraction of afterglows from BNS mergers with flux larger than a certain value (displayed on the $y$-axis), as a function of time from the merger, in three different observation bands: radio, optical and X-rays, from left to right. 
For each flux, the probabilities are separately displayed for BNS merger events in the four mass bins of our study. 
While Fig.~\ref{fig:aft_ran} assumes a random distribution for the viewing angle with respect to the emitting jet, Fig.~\ref{fig:aft_obs} has been computed with the viewing angles drawn from the probability distribution in Eq.~(\ref{eq:pobs}). The contour lines in both figures represent the fraction of simulated events with flux above the corresponding values on the $y$ axis, at the observing times (from the time of merger) indicated in the $x$ axis. A common feature among all panels is that bright events only happen at earlier times and with small probability, with the dimmest, late-peaking events being more probable. This is due to the fact that both increasing the viewing angle and decreasing the interstellar density results in less bright events that peak at late times (weeks to months). The prevalence of intermediate densities (Fig.~\ref{fig:densities}) and the geometry that favors large viewing angles results in rare, early-peaking, bright events and more common, late-peaking dim events. Events peaking at more than a few months are instead extremely rare because the viewing angle cannot exceed 90$^\circ$ and binaries merging in extremely low densities are more rare. (Fig.~\ref{fig:densities}).

A visual comparison between Fig.~\ref{fig:aft_ran} and Fig.~\ref{fig:aft_obs} shows that the latter has a larger fraction of brighter events, and that the time from the merger at which the events reach their maximum brightness is typically smaller. This results from the fact that the events in Fig.~\ref{fig:aft_obs} are selected with a bias towards smaller viewing angles with respect to the completely random selection of the events in Fig.~\ref{fig:aft_ran}. 
Smaller viewing angles imply both brighter emission as well as earlier peak emission.

The figures also indicate with an horizontal line representative threshold detections of current instruments, that is {\em Chandra} and {\em Swift} in X-rays, HST and Robotic follow up for the optical, and a 1hr integration time on the VLA for the radio.
These instruments were among the ones which detected the afterglow emission of GRB170817A. The results of Fig.~\ref{fig:aft_obs} imply that an event like GRB170817A, which was initially triggered in GWs and which had a viewing angle inferred from EM emission consistent with that from GWs, had a probability $\gtrsim 0.5$ of being observed in all afterglow wavelength bands (recall the the redshift of $z=0.01$ corresponds to a distance of 45~Mpc, close to that of GW/GRB170817A). 

Next we examine the dependence on the galaxy mass, studied by means of the four rows in Fig.~\ref{fig:aft_ran} and Fig.~\ref{fig:aft_obs}. In both cases, there is a clear bias against detection of events from smaller galaxies. The exact value varies depending on the band, and is slightly different between the two choices of viewing angles. A quantitative comparison between the peak emission from events in the least massive galaxy group (M1) and the most massive one (M4) shows that, for example, in the radio band, there is an enhanced probability by a factor of $\sim 2.5:1$ of detecting an event from a massive galaxy than one from a smaller one (assuming the intrinsic number of events being the same). Therefore, in addition to being disfavored by event number (cfr. Fig.~\ref{fig:fBNS}), BNS mergers in small galaxies suffer by a selection bias, which makes their detectable fraction suppressed compared to the intrinsic value. 

For the higher redshits snapshots $z=0.1$ and $z=1$, which are beyond the 
current LIGO--Virgo horizon (as a reference, $z=0.1$ corresponds to a luminosity distance of about 480 Mpc), we simulate the statistical properties of the EM counterparts from BNS mergers
as for the standard cosmological short GRBs. More specifically, we consider a random distribution on the sky of the viewing angle $\theta_{\rm view}$ with respect to the observer line of sight. 
We then compute the $\gamma$-ray flux in the {\em Swift/XRT} and {\em Fermi/GBM} bands. As observed above, the threshold sensitivites are roughly comparable for these two instruments. 
If the flux is above the detection limit, then we compute its afterglow based on the merger location of the BNS within its corresponding host galaxies.
Given the  anti-correlation between flux and viewing angle (cfr. Fig.~\ref{fig:prompt}), this procedure is practically equivalent to selecting the maximum viewing angle,  $\theta_{\rm view,\gamma}$, which can allow detection at the given redshift. We find $\theta_{\rm view,\gamma}\sim 29.4$~deg for merger events at  $z=0.1$, and $\theta_{\rm view,\gamma}\sim 11.5$~deg for those at the redshift $z=1$.

The detection probabilities for the afterglows of the SGRBs triggered in $\gamma$-rays are given in Fig.~\ref{fig:aftz0.1} and Fig.~\ref{fig:aftz1} for the two redshifts $z=0.1$ and $z=1$, respectively. The general trends with galaxy mass are similar to those found at $z=0.01$, and that is that low-mass galaxies are selectively disfavoured for afterglow detection. Since this is the key to localize the burst and hence measure its redshift via host galaxy identification, we conclude that the population of SGRBs does not represent an unbiased distribution with respect to the underlying one with respect to the galaxy host. This needs to be kept in mind when comparing theoretical models of SGRBs from BNSs with the statistical properties of their host galaxies. 

If the afterglow properties of GRB170717 are indeed representative of the bulk of the cosmological SGRBs, then our simulations predict that at least half of the events should be detectable with current X-ray detectors and an optical telescopes such as the Hubble. In the radio, an hour of integration time with the VLA yields $\lesssim 10\%$ of detectable events.

Generally speaking, the relative observability in different bands depends on both the viewing angle and the ambient density. The latter is a more important factor for the radio band than the X-ray band at the times at which the emitting flow is in the so-called synchrotron radiative regime, i.e. a regime in which radiative losses are significant. This regime, which is satisfied at emitting frequencies above a critical frequency value, is hence more likely satisfied for the higher energy bands than the lower one. Once in this regime, the emission becomes independent of the ambient density (\citealt{Sari1998}, see also discussion in \citealt{Saleem2018a}). 

A noticeable difference between the population of GW-detected afterglow counterparts, and the population of $\gamma$-ray triggered afterglow counterparts at the higher redshift snapshots $z=0.1$ and $z=1$, is the fact that in the latter cases, the relative fraction of X-ray to radio (and optical) detectable counterparts is significantly higher than for the former. This is a direct consequence of the fact that $\gamma$-ray triggered events are generally selecting out at much smaller viewing angles than GW-detected events. As shown by Fig.~\ref{fig:luminosities} and discussed in the corresponding text, at smaller viewing angles the X-ray/radio relative flux is much larger than it is when the event is observed from large viewing angles. 
For this reason, since the maximum viewing angle $\theta_{\rm view,\gamma}$ (to trigger bursts in $\gamma$-rays) is smaller for the $z=1$ events than it is for the $z=0.1$ ones, the tendency for the X-ray flux to be brighter than the radio one is further enhanced in the 
higher redshift snapshot than it is in the $z=0.1$ one.

We note that in the Monte Carlo simulations leading to Figs.~\ref{fig:aft_obs} through Fig.~\ref{fig:aftz1} we did not include the contribution from the kilonova, which was observed for GW170817 \citep{abbottmultimessenger,coulter2017,Kasen2017,cowperthwaite2017,pian2017,Smartt2017,Drout2017,Tanvir2017,soares-santos2017,Arcavi2017,chornock2017}. The maximum flux in the optical (red) was around 200~$\mu$~Jy at about 1~day. At the redshift of $z=0.01$ (roughly the distance to GW170817), this luminosity is comparable to the brightest afterglows in Fig.~\ref{fig:aft_ran} and ~\ref{fig:aft_obs}. More specifically, the kilonova luminosity exceeds the maximum afterglow luminosity (in the red band used here) for viewing angles $\theta_{\rm view} \gtrsim 10$~deg  for an ambient density $n_{\rm ISM}=0.01$~cm$^{-3}$ (higher densities correlate with larger values of $\theta_{\rm view}$ for the kilonova and afterglow luminosities to be comparable, and viceversa).

We further note that the afterglow calculations have not included absorption by line of sight material within the host galaxy. This might affect optical and X-ray detectability due to dust absorption and photoionization, respectively. Modelling this effect would require assumption on metallicity and dust to gas ratios that are uncertain and beyond the scope of this work.

\section{Summary and Discussion}

The association of GW170817 with GRB170817A, and the recognition that the latter has properties fully consistent with those of the standard cosmological short GRBs, has opened a new line of investigation of short GRBs, and in particular their detectability in connection with GW-triggered BNS mergers. The observed variety of SGRB luminosities is consistent with being largely driven by viewing angle effects, with GRB170817A being a rather typical event among the well studied set of cosmological SGRBs \citep{Wu2019}. 

In this work we have performed a comprehensive study of the EM counterparts expected from BNS merger events, using the intrinsic source properties of GRB170817A as a template. Our population of BNSs is generated via population synthesis calculations with the code {\sc mobse}, and seeded in a sample of galaxies from the TNG50 simulation, at three representative redshifts of $z=0.01, 0.1, 1$, straddling the range of GW-detected events and that of the standard cosmological short GRBs, detected in $\gamma$-rays.

We have studied the BNS population by dividing their host galaxies from the TNG50 simulation in 4 mass groups at each redshift snapshot, with the goal of uncovering possible biases of the observed population with respect to the intrinsic one. Our analysis uncovered that such biases do indeed exist.
 
 Additionally, the comparative analysis of the afterglow counterparts to GW-triggered events, versus $\gamma$-ray triggered events (i.e. the standard cosmological GRBs) has uncovered some interesting differences among the two populations, even as the underlying source is the same. 

Our main results are summarized in the following:

GRB170817A, first detected in GWs and later followed up in the electromagnetic spectrum, appears to be a rather common/typical event for the BNS merging population. Our population synthesis calculations of BNSs at the redshift snapshot $z=0.01$ of the TNG50 simulation, which, corresponding to a distance of $\sim 45$~Mpc, is very close to that of GW170817  (40~Mpc), shows that the broadband detection of GRB170817A did not require especially lucky circumstances\footnote{We remark that this statement specifically refers to the probability of observing the broadband EM counterpart to GW170817. On the other hand, the host galaxy of this event appears somewhat unusual \citep{Palmese2017}.}. In $\gamma$-rays, the {\em Fermi GBM} detector would have $\sim 30\%$ probability to detect such an event if its EM emitting jet were uncorrelated with the orbital plane of the merging NSs (having corrected the probability of Fig.~\ref{fig:fluxes-gamma} by the field of view of 9.5~sr of this telescope), and of $\sim 60\%$ for the situation in which the relativistic jet is roughly aligned with the orbital angular momentum of the binary, as found to be the case for this source.

Our calculations of the distribution of the merger sites of BNSs within the sample of galaxies from the TNG50 simulation has allowed us to predict the distribution of interstellar medium densities, and thus the afterglow brightness as a function of time and wavelength.  At the redshift of $z=0.01$, about 30\% of BNS mergers are expected to have detectable afterglow radiation for randomly oriented jets, whereas the fraction is enhanced to $\gtrsim 50\%$ for jets aligned with the orbital angular momentum of the binary. 
Therefore, the binary NS merger event GRB170817A appears to be representative of the theoretically predicted  population of BNS mergers in the local Universe.

To date, there has been only one other reported GW-detected BNS merger event,
GW190425 \citep{abbottGW190425}, at a distance of about 155~Mpc. Both the {\em Fermi/BAT} \citep{FermiGCN190425} and the {\rm Swift/XRT} telescopes \citep{SwiftGCN190425} were located in unfavorable positions for observability, with $\sim 45\%$ of the GW localization area located behind the Earth. This, combined with the large distance of the event prevented a precise localization of the source, hampering follow-up from ground based observatories. Several candidate transients were detected, but were all eventually discarded as counterpart to the BNS merger
\citep{Song2019,Paterson2021}.

 Our study of the theoretically predicted properties of the BNS population in their host galaxies shows that there is a significant bias towards detecting events in small galaxies with respect to the larger ones. More specifically, for nearby BNS mergers which are GW-triggered, the probability of detecting an afterglow from a galaxy in the interval range $8 < \log[{\cal M_{\rm gal,1}}/{\rm M}_\odot ]< 8.75$ is suppressed by about a factor of two with respect to BNS mergers in galaxies of mass within $10.25 < \log[{\cal M_{\rm gal,1}}/{\rm M}_\odot ]< 11$. 
Since afterglow detections are key to host identification, this bias needs to be accounted for when extracting physical information on the underlying distribution of BNS mergers.

In addition to the snapshot at $z=0.01$, which is well within the LIGO-Virgo current horizon, we studied the properties of the BNS population and its detectable counterparts also at the two higher redshift snapshots from TNG50 of $z=0.1$ and $z=1$. BNSs from a redshift distance $z=1$ are beyond detection by current GW detectors, even at design sensitivity, but are expected to be detectable with the future Einstein Telescope \citep{Punturo2010} and the Cosmic Explorer \citep{Dwyer2015}. On the other hand, BNS mergers occurring at $z\sim 0.1$, while beyond the horizon of the current instruments, will be within reach of LIGO and Virgo at design sensitivities
\footnote{https://dcc.ligo.org/LIGO-T1800133/public}, and even more so with the addition of KAGRA and LIGO-India (\citealt{Nissanke2013}; for a comparison among detectors see e.g. Figure~1 of \citealt{Yu2021}). 

For BNS mergers which are outside of the detectable GW horizon, the trigger is in $\gamma$-rays, and that is then followed up at longer wavelengths. These are the standard cosmological short GRBs (or at least a fraction of them, if a contribution is given also by NS-BH mergers). For a $\gamma$-ray detection with the {\em Swift XRT} and {\em Fermi GBM} telescopes, we find that a GRB170817A-like event would need to be observed within a viewing angle from the jet axis of $\sim 29$ deg  at $z=0.1$, and of  $\sim 11$~deg at $z=1$.  At these smaller viewing angles, the afterglow brightness ratio  X-ray/O, but especially X-rays/R  is larger than it is at larger viewing angles. As a result, a relatively larger fraction (compared with the GW-detected events) will be seen in X-rays than in Optical, and even more so in Radio. 

It is interesting to compare our theoretical results with observations of cosmological short GRBs. As summarized in the review by
\citet{berger2014}, for short GRBs, the broadest and most homogeneous data set is in the X-ray band from the {\em Swift/XRT} satellite, with about 50 X-ray  afterglow detections. Of these, about half have also an optical afterglow detection, and only a handful do so also in radio. 
The broad consistency between our theoretical predictions and the broadband observations of the cosmological short GRBs, having taken the jet properties of GRB170817A as 'canonical',  provide yet another piece of evidence that indeed the bulk of the cosmological short GRB population is produced by GRB170817A-like events.

Looking into the future, while here we have made the first step in combining state-of-the art population synthesis calculations, galaxy simulations and numerical broadband light curves to theoretically predict the observable properties of the short GRB population from GRB170817A-like BNS mergers events, there are several natural extensions of our study which we plan to address in future work. First is to consider a range in the intrinsic microphysics jet parameters. Since GRB170817A appears to be an average event compared to the bulk of short GRBs \citep{Wu2019}, we do not expect any quantitative change in the results, but there will be a spread in the luminosity brightness distributions. Most importantly, it will be interesting to compare statistically the EM properties of a population of field binary NSs with the properties of dynamically formed binary NSs, to see whether there are any telltale features which can help discriminate the two formation channels. Also importantly, it will be  useful a comparison with the EM properties of an NS-BH merging population (albeit to date we are lacking an observationally-derived prototype light curve) to help assess the fraction of short GRBs (if any) which is due to this interesting formation channel.

\section*{Acknowledgements}
 RP and YW gratefully acknowledge support by NSF award AST-2006839.
MCA acknowledges financial support from the Austrian National Science Foundation through FWF stand-alone grant P31154-N27. MM and FS acknowledge financial support from the European Research Council for the ERC Consolidator grant DEMOBLACK, under contract no. 770017. DL acknowledges suport from NASA grant NNX17AK42G (ATP) and NSF grant AST-1907955.

\section*{Data Availability}

All the simulation data produced for this paper will be made available upon request. 
 



\bibliographystyle{mnras}
\bibliography{bibio} 

\begin{thebibliography}{}
\makeatletter
\relax
\def\mn@urlcharsother{\let\do\@makeother \do\$\do\&\do\#\do\^\do\_\do\%\do\~}
\def\mn@doi{\begingroup\mn@urlcharsother \@ifnextchar [ {\mn@doi@}
  {\mn@doi@[]}}
\def\mn@doi@[#1]#2{\def\@tempa{#1}\ifx\@tempa\@empty \href
  {http://dx.doi.org/#2} {doi:#2}\else \href {http://dx.doi.org/#2} {#1}\fi
  \endgroup}
\def\mn@eprint#1#2{\mn@eprint@#1:#2::\@nil}
\def\mn@eprint@arXiv#1{\href {http://arxiv.org/abs/#1} {{\tt arXiv:#1}}}
\def\mn@eprint@dblp#1{\href {http://dblp.uni-trier.de/rec/bibtex/#1.xml}
  {dblp:#1}}
\def\mn@eprint@#1:#2:#3:#4\@nil{\def\@tempa {#1}\def\@tempb {#2}\def\@tempc
  {#3}\ifx \@tempc \@empty \let \@tempc \@tempb \let \@tempb \@tempa \fi \ifx
  \@tempb \@empty \def\@tempb {arXiv}\fi \@ifundefined
  {mn@eprint@\@tempb}{\@tempb:\@tempc}{\expandafter \expandafter \csname
  mn@eprint@\@tempb\endcsname \expandafter{\@tempc}}}

\bibitem[\protect\citeauthoryear{{Abbott} et~al.,}{{Abbott}
  et~al.}{2016}]{abbott2018observingscenario}
{Abbott} B.~P.,  et~al., 2016, \mn@doi [Living Reviews in Relativity]
  {10.1007/lrr-2016-1}, \href
  {https://ui.adsabs.harvard.edu/abs/2016LRR....19....1A} {19, 1}

\bibitem[\protect\citeauthoryear{{Abbott} et~al.,}{{Abbott}
  et~al.}{2017a}]{abbottGW170817}
{Abbott} B.~P.,  et~al., 2017a, \mn@doi [Physical Review Letters]
  {10.1103/PhysRevLett.119.161101}, \href
  {http://adsabs.harvard.edu/abs/2017PhRvL.119p1101A} {119, 161101}

\bibitem[\protect\citeauthoryear{{Abbott} et~al.,}{{Abbott}
  et~al.}{2017b}]{abbottmultimessenger}
{Abbott} B.~P.,  et~al., 2017b, \mn@doi [\apjl] {10.3847/2041-8213/aa91c9},
  \href {http://adsabs.harvard.edu/abs/2017ApJ...848L..12A} {848, L12}

\bibitem[\protect\citeauthoryear{{Abbott} et~al.,}{{Abbott}
  et~al.}{2020}]{abbottGW190425}
{Abbott} B.~P.,  et~al., 2020, \mn@doi [\apjl] {10.3847/2041-8213/ab75f5},
  \href {https://ui.adsabs.harvard.edu/abs/2020ApJ...892L...3A} {892, L3}

\bibitem[\protect\citeauthoryear{{Adhikari}, {Fishbach}, {Holz}, {Wechsler}  \&
  {Fang}}{{Adhikari} et~al.}{2020}]{Adhikari2020}
{Adhikari} S.,  {Fishbach} M.,  {Holz} D.~E.,  {Wechsler} R.~H.,   {Fang} Z.,
  2020, \mn@doi [\apj] {10.3847/1538-4357/abbfb7}, \href
  {https://ui.adsabs.harvard.edu/abs/2020ApJ...905...21A} {905, 21}

\bibitem[\protect\citeauthoryear{{Alexander} et~al.,}{{Alexander}
  et~al.}{2017}]{Alexander2017}
{Alexander} K.~D.,  et~al., 2017, \mn@doi [\apjl] {10.3847/2041-8213/aa905d},
  \href {https://ui.adsabs.harvard.edu/abs/2017ApJ...848L..21A} {848, L21}

\bibitem[\protect\citeauthoryear{{Alexander} et~al.,}{{Alexander}
  et~al.}{2018}]{Alexander2018}
{Alexander} K.~D.,  et~al., 2018, \mn@doi [\apjl] {10.3847/2041-8213/aad637},
  \href {https://ui.adsabs.harvard.edu/abs/2018ApJ...863L..18A} {863, L18}

\bibitem[\protect\citeauthoryear{{Antonelli} et~al.,}{{Antonelli}
  et~al.}{2009}]{Antonelli2009}
{Antonelli} L.~A.,  et~al., 2009, \mn@doi [\aap] {10.1051/0004-6361/200913062},
  \href {https://ui.adsabs.harvard.edu/abs/2009A&A...507L..45A} {507, L45}

\bibitem[\protect\citeauthoryear{{Antonini} \& {Gieles}}{{Antonini} \&
  {Gieles}}{2020}]{antonini2020}
{Antonini} F.,  {Gieles} M.,  2020, \mn@doi [\mnras] {10.1093/mnras/stz3584},
  \href {https://ui.adsabs.harvard.edu/abs/2020MNRAS.492.2936A} {492, 2936}

\bibitem[\protect\citeauthoryear{{Antonini} \& {Rasio}}{{Antonini} \&
  {Rasio}}{2016}]{antonini2016}
{Antonini} F.,  {Rasio} F.~A.,  2016, \mn@doi [\apj]
  {10.3847/0004-637X/831/2/187}, \href
  {http://adsabs.harvard.edu/abs/2016ApJ...831..187A} {831, 187}

\bibitem[\protect\citeauthoryear{{Antonini}, {Chatterjee}, {Rodriguez},
  {Morscher}, {Pattabiraman}, {Kalogera}  \& {Rasio}}{{Antonini}
  et~al.}{2016}]{Antonini2016b}
{Antonini} F.,  {Chatterjee} S.,  {Rodriguez} C.~L.,  {Morscher} M.,
  {Pattabiraman} B.,  {Kalogera} V.,   {Rasio} F.~A.,  2016, \mn@doi [\apj]
  {10.3847/0004-637X/816/2/65}, \href
  {https://ui.adsabs.harvard.edu/abs/2016ApJ...816...65A} {816, 65}

\bibitem[\protect\citeauthoryear{{Arca Sedda}}{{Arca
  Sedda}}{2020}]{arcasedda2020b}
{Arca Sedda} M.,  2020, \mn@doi [\apj] {10.3847/1538-4357/ab723b}, \href
  {https://ui.adsabs.harvard.edu/abs/2020ApJ...891...47A} {891, 47}

\bibitem[\protect\citeauthoryear{{Arca Sedda}, {Mapelli}, {Spera},
  {Benacquista}  \& {Giacobbo}}{{Arca Sedda} et~al.}{2020}]{arcasedda2020}
{Arca Sedda} M.,  {Mapelli} M.,  {Spera} M.,  {Benacquista} M.,   {Giacobbo}
  N.,  2020, \mn@doi [\apj] {10.3847/1538-4357/ab88b2}, \href
  {https://ui.adsabs.harvard.edu/abs/2020ApJ...894..133A} {894, 133}

\bibitem[\protect\citeauthoryear{{Arcavi} et~al.,}{{Arcavi}
  et~al.}{2017}]{Arcavi2017}
{Arcavi} I.,  et~al., 2017, \mn@doi [\nat] {10.1038/nature24291}, \href
  {https://ui.adsabs.harvard.edu/abs/2017Natur.551...64A} {551, 64}

\bibitem[\protect\citeauthoryear{{Artale}, {Mapelli}, {Giacobbo}, {Sabha},
  {Spera}, {Santoliquido}  \& {Bressan}}{{Artale} et~al.}{2019}]{Artale2019}
{Artale} M.~C.,  {Mapelli} M.,  {Giacobbo} N.,  {Sabha} N.~B.,  {Spera} M.,
  {Santoliquido} F.,   {Bressan} A.,  2019, \mn@doi [\mnras]
  {10.1093/mnras/stz1382}, \href
  {https://ui.adsabs.harvard.edu/abs/2019MNRAS.487.1675A} {487, 1675}

\bibitem[\protect\citeauthoryear{{Artale}, {Mapelli}, {Bouffanais}, {Giacobbo},
  {Pasquato}  \& {Spera}}{{Artale} et~al.}{2020a}]{Artale2020}
{Artale} M.~C.,  {Mapelli} M.,  {Bouffanais} Y.,  {Giacobbo} N.,  {Pasquato}
  M.,   {Spera} M.,  2020a, \mn@doi [\mnras] {10.1093/mnras/stz3190}, \href
  {https://ui.adsabs.harvard.edu/abs/2020MNRAS.491.3419A} {491, 3419}

\bibitem[\protect\citeauthoryear{{Artale}, {Bouffanais}, {Mapelli}, {Giacobbo},
  {Sabha}, {Santoliquido}, {Pasquato}  \& {Spera}}{{Artale}
  et~al.}{2020b}]{Artale2020b}
{Artale} M.~C.,  {Bouffanais} Y.,  {Mapelli} M.,  {Giacobbo} N.,  {Sabha}
  N.~B.,  {Santoliquido} F.,  {Pasquato} M.,   {Spera} M.,  2020b, \mn@doi
  [\mnras] {10.1093/mnras/staa1252}, \href
  {https://ui.adsabs.harvard.edu/abs/2020MNRAS.495.1841A} {495, 1841}

\bibitem[\protect\citeauthoryear{{Askar}, {Szkudlarek}, {Gondek-Rosi{\'n}ska},
  {Giersz}  \& {Bulik}}{{Askar} et~al.}{2017}]{askar2017}
{Askar} A.,  {Szkudlarek} M.,  {Gondek-Rosi{\'n}ska} D.,  {Giersz} M.,
  {Bulik} T.,  2017, \mn@doi [\mnras] {10.1093/mnrasl/slw177}, \href
  {http://adsabs.harvard.edu/abs/2017MNRAS.464L..36A} {464, L36}

\bibitem[\protect\citeauthoryear{{Band} et~al.,}{{Band}
  et~al.}{1993}]{Band1993}
{Band} D.,  et~al., 1993, \mn@doi [\apj] {10.1086/172995}, \href
  {https://ui.adsabs.harvard.edu/abs/1993ApJ...413..281B} {413, 281}

\bibitem[\protect\citeauthoryear{{Banerjee}}{{Banerjee}}{2017}]{banerjee2017}
{Banerjee} S.,  2017, \mn@doi [\mnras] {10.1093/mnras/stw3392}, \href
  {http://adsabs.harvard.edu/abs/2017MNRAS.467..524B} {467, 524}

\bibitem[\protect\citeauthoryear{{Banerjee}}{{Banerjee}}{2020}]{banerjee2020}
{Banerjee} S.,  2020, arXiv e-prints, \href
  {https://ui.adsabs.harvard.edu/abs/2020arXiv200407382B} {p. arXiv:2004.07382}

\bibitem[\protect\citeauthoryear{{Banerjee}, {Baumgardt}  \&
  {Kroupa}}{{Banerjee} et~al.}{2010}]{banerjee2010}
{Banerjee} S.,  {Baumgardt} H.,   {Kroupa} P.,  2010, \mn@doi [\mnras]
  {10.1111/j.1365-2966.2009.15880.x}, \href
  {http://adsabs.harvard.edu/abs/2010MNRAS.402..371B} {402, 371}

\bibitem[\protect\citeauthoryear{{Bartos}, {Kocsis}, {Haiman}  \&
  {M{\'a}rka}}{{Bartos} et~al.}{2017}]{bartos2017}
{Bartos} I.,  {Kocsis} B.,  {Haiman} Z.,   {M{\'a}rka} S.,  2017, \mn@doi
  [\apj] {10.3847/1538-4357/835/2/165}, \href
  {https://ui.adsabs.harvard.edu/abs/2017ApJ...835..165B} {835, 165}

\bibitem[\protect\citeauthoryear{{Belczynski}, {Kalogera}  \&
  {Bulik}}{{Belczynski} et~al.}{2002}]{belczynski2002}
{Belczynski} K.,  {Kalogera} V.,   {Bulik} T.,  2002, \mn@doi [\apj]
  {10.1086/340304}, \href {http://adsabs.harvard.edu/abs/2002ApJ...572..407B}
  {572, 407}

\bibitem[\protect\citeauthoryear{{Belczynski}, {Perna}, {Bulik}, {Kalogera},
  {Ivanova}  \& {Lamb}}{{Belczynski} et~al.}{2006}]{Belczynski2006gal}
{Belczynski} K.,  {Perna} R.,  {Bulik} T.,  {Kalogera} V.,  {Ivanova} N.,
  {Lamb} D.~Q.,  2006, \mn@doi [\apj] {10.1086/505169}, \href
  {https://ui.adsabs.harvard.edu/abs/2006ApJ...648.1110B} {648, 1110}

\bibitem[\protect\citeauthoryear{{Belczynski}, {Taam}, {Kalogera}, {Rasio}  \&
  {Bulik}}{{Belczynski} et~al.}{2007}]{belczynski2007}
{Belczynski} K.,  {Taam} R.~E.,  {Kalogera} V.,  {Rasio} F.~A.,   {Bulik} T.,
  2007, \mn@doi [\apj] {10.1086/513562}, \href
  {http://adsabs.harvard.edu/abs/2007ApJ...662..504B} {662, 504}

\bibitem[\protect\citeauthoryear{{Belczynski}, {Holz}, {Bulik}  \&
  {O'Shaughnessy}}{{Belczynski} et~al.}{2016}]{belczynski2016}
{Belczynski} K.,  {Holz} D.~E.,  {Bulik} T.,   {O'Shaughnessy} R.,  2016,
  \mn@doi [\nat] {10.1038/nature18322}, \href
  {http://adsabs.harvard.edu/abs/2016Natur.534..512B} {534, 512}

\bibitem[\protect\citeauthoryear{{Belczynski}, {Ryu}, {Perna}, {Berti},
  {Tanaka}  \& {Bulik}}{{Belczynski} et~al.}{2017}]{Belczynski2017}
{Belczynski} K.,  {Ryu} T.,  {Perna} R.,  {Berti} E.,  {Tanaka} T.~L.,
  {Bulik} T.,  2017, \mn@doi [\mnras] {10.1093/mnras/stx1759}, \href
  {https://ui.adsabs.harvard.edu/abs/2017MNRAS.471.4702B} {471, 4702}

\bibitem[\protect\citeauthoryear{{Belczynski} et~al.,}{{Belczynski}
  et~al.}{2018}]{Belczynski2018}
{Belczynski} K.,  et~al., 2018, \mn@doi [\aap] {10.1051/0004-6361/201732428},
  \href {https://ui.adsabs.harvard.edu/abs/2018A&A...615A..91B} {615, A91}

\bibitem[\protect\citeauthoryear{{Berger}}{{Berger}}{2010}]{Berger2010}
{Berger} E.,  2010, \mn@doi [\apj] {10.1088/0004-637X/722/2/1946}, \href
  {https://ui.adsabs.harvard.edu/abs/2010ApJ...722.1946B} {722, 1946}

\bibitem[\protect\citeauthoryear{{Berger}}{{Berger}}{2014}]{berger2014}
{Berger} E.,  2014, \mn@doi [\araa] {10.1146/annurev-astro-081913-035926},
  \href {http://adsabs.harvard.edu/abs/2014ARA%26A..52...43B} {52, 43}

\bibitem[\protect\citeauthoryear{{Bhat} et~al.,}{{Bhat}
  et~al.}{2016}]{Bhat2016}
{Bhat} P.~N.,  et~al., 2016, VizieR Online Data Catalog, \href
  {https://ui.adsabs.harvard.edu/abs/2016yCat..22230028B} {p. J/ApJS/223/28}

\bibitem[\protect\citeauthoryear{{Bloom}}{{Bloom}}{2003}]{Bloom2003}
{Bloom} J.~S.,  2003, \mn@doi [\pasp] {10.1086/345917}, \href
  {https://ui.adsabs.harvard.edu/abs/2003PASP..115..271B} {115, 271}

\bibitem[\protect\citeauthoryear{{Briel}, {Eldridge}, {Stanway}, {Stevance}  \&
  {Chrimes}}{{Briel} et~al.}{2021}]{Briel2021}
{Briel} M.~M.,  {Eldridge} J.~J.,  {Stanway} E.~R.,  {Stevance} H.~F.,
  {Chrimes} A.~A.,  2021, arXiv e-prints, \href
  {https://ui.adsabs.harvard.edu/abs/2021arXiv211108124B} {p. arXiv:2111.08124}

\bibitem[\protect\citeauthoryear{{Camelio}, {Dietrich}, {Rosswog}  \&
  {Haskell}}{{Camelio} et~al.}{2021}]{Camelio2021}
{Camelio} G.,  {Dietrich} T.,  {Rosswog} S.,   {Haskell} B.,  2021, \mn@doi
  [\prd] {10.1103/PhysRevD.103.063014}, \href
  {https://ui.adsabs.harvard.edu/abs/2021PhRvD.103f3014C} {103, 063014}

\bibitem[\protect\citeauthoryear{{Cao}, {Lu}  \& {Zhao}}{{Cao}
  et~al.}{2018}]{Cao2018}
{Cao} L.,  {Lu} Y.,   {Zhao} Y.,  2018, \mn@doi [\mnras]
  {10.1093/mnras/stx3087}, \href
  {https://ui.adsabs.harvard.edu/abs/2018MNRAS.474.4997C} {474, 4997}

\bibitem[\protect\citeauthoryear{Chornock et~al.,}{Chornock
  et~al.}{2017}]{chornock2017}
Chornock R.,  et~al., 2017, The Astrophysical Journal Letters, 848, L19

\bibitem[\protect\citeauthoryear{{Chruslinska}, {Belczynski}, {Klencki}  \&
  {Benacquista}}{{Chruslinska} et~al.}{2018}]{chruslinska2018}
{Chruslinska} M.,  {Belczynski} K.,  {Klencki} J.,   {Benacquista} M.,  2018,
  \mn@doi [\mnras] {10.1093/mnras/stx2923}, \href
  {http://adsabs.harvard.edu/abs/2018MNRAS.474.2937C} {474, 2937}

\bibitem[\protect\citeauthoryear{{Chruslinska}, {Nelemans}  \&
  {Belczynski}}{{Chruslinska} et~al.}{2019}]{chruslinska2019}
{Chruslinska} M.,  {Nelemans} G.,   {Belczynski} K.,  2019, \mn@doi [\mnras]
  {10.1093/mnras/sty3087}, \href
  {https://ui.adsabs.harvard.edu/abs/2019MNRAS.482.5012C} {482, 5012}

\bibitem[\protect\citeauthoryear{{Chu}, {Yu}  \& {Lu}}{{Chu}
  et~al.}{2021}]{Chu2021}
{Chu} Q.,  {Yu} S.,   {Lu} Y.,  2021, \mn@doi [\mnras]
  {10.1093/mnras/stab2882}, \href
  {https://ui.adsabs.harvard.edu/abs/2021MNRAS.tmp.2657C} {}

\bibitem[\protect\citeauthoryear{{Ciolfi}, {Kastaun}, {Giacomazzo}, {Endrizzi},
  {Siegel}  \& {Perna}}{{Ciolfi} et~al.}{2017}]{Ciolfi2017}
{Ciolfi} R.,  {Kastaun} W.,  {Giacomazzo} B.,  {Endrizzi} A.,  {Siegel} D.~M.,
   {Perna} R.,  2017, \mn@doi [\prd] {10.1103/PhysRevD.95.063016}, \href
  {https://ui.adsabs.harvard.edu/abs/2017PhRvD..95f3016C} {95, 063016}

\bibitem[\protect\citeauthoryear{{Coulter} et~al.,}{{Coulter}
  et~al.}{2017}]{coulter2017}
{Coulter} D.~A.,  et~al., 2017, \mn@doi [Science] {10.1126/science.aap9811},
  \href {http://adsabs.harvard.edu/abs/2017Sci...358.1556C} {358, 1556}

\bibitem[\protect\citeauthoryear{{Covino} et~al.,}{{Covino}
  et~al.}{2006}]{Covino2006}
{Covino} S.,  et~al., 2006, \mn@doi [\aap] {10.1051/0004-6361:200500228}, \href
  {https://ui.adsabs.harvard.edu/abs/2006A&A...447L...5C} {447, L5}

\bibitem[\protect\citeauthoryear{Cowperthwaite et~al.,}{Cowperthwaite
  et~al.}{2017}]{cowperthwaite2017}
Cowperthwaite P.~S.,  et~al., 2017, The Astrophysical Journal Letters, 848, L17

\bibitem[\protect\citeauthoryear{{D'Avanzo} et~al.,}{{D'Avanzo}
  et~al.}{2009}]{Davanzo2009}
{D'Avanzo} P.,  et~al., 2009, \mn@doi [\aap] {10.1051/0004-6361/200811294},
  \href {https://ui.adsabs.harvard.edu/abs/2009A&A...498..711D} {498, 711}

\bibitem[\protect\citeauthoryear{{Di Carlo}, {Mapelli}, {Bouffanais},
  {Giacobbo}, {Bressan}, {Spera}  \& {Haardt}}{{Di Carlo}
  et~al.}{2019a}]{dicarlo2019b}
{Di Carlo} U.~N.,  {Mapelli} M.,  {Bouffanais} Y.,  {Giacobbo} N.,  {Bressan}
  S.,  {Spera} M.,   {Haardt} F.,  2019a, arXiv e-prints, \href
  {https://ui.adsabs.harvard.edu/abs/2019arXiv191101434D} {p. arXiv:1911.01434}

\bibitem[\protect\citeauthoryear{{Di Carlo}, {Giacobbo}, {Mapelli}, {Pasquato},
  {Spera}, {Wang}  \& {Haardt}}{{Di Carlo} et~al.}{2019b}]{dicarlo2019a}
{Di Carlo} U.~N.,  {Giacobbo} N.,  {Mapelli} M.,  {Pasquato} M.,  {Spera} M.,
  {Wang} L.,   {Haardt} F.,  2019b, \mn@doi [\mnras] {10.1093/mnras/stz1453},
  \href {https://ui.adsabs.harvard.edu/abs/2019MNRAS.487.2947D} {487, 2947}

\bibitem[\protect\citeauthoryear{{Di Carlo} et~al.,}{{Di Carlo}
  et~al.}{2020}]{dicarlo2020}
{Di Carlo} U.~N.,  et~al., 2020, arXiv e-prints, \href
  {https://ui.adsabs.harvard.edu/abs/2020arXiv200409525D} {p. arXiv:2004.09525}

\bibitem[\protect\citeauthoryear{{Dobie} et~al.,}{{Dobie}
  et~al.}{2018}]{Dobie2018}
{Dobie} D.,  et~al., 2018, \mn@doi [\apjl] {10.3847/2041-8213/aac105}, \href
  {https://ui.adsabs.harvard.edu/abs/2018ApJ...858L..15D} {858, L15}

\bibitem[\protect\citeauthoryear{{Dominik}, {Belczynski}, {Fryer}, {Holz},
  {Berti}, {Bulik}, {Mandel}  \& {O'Shaughnessy}}{{Dominik}
  et~al.}{2012}]{dominik2012}
{Dominik} M.,  {Belczynski} K.,  {Fryer} C.,  {Holz} D.~E.,  {Berti} E.,
  {Bulik} T.,  {Mandel} I.,   {O'Shaughnessy} R.,  2012, \mn@doi [\apj]
  {10.1088/0004-637X/759/1/52}, \href
  {http://adsabs.harvard.edu/abs/2012ApJ...759...52D} {759, 52}

\bibitem[\protect\citeauthoryear{{Dominik}, {Belczynski}, {Fryer}, {Holz},
  {Berti}, {Bulik}, {Mandel}  \& {O'Shaughnessy}}{{Dominik}
  et~al.}{2013}]{dominik2013}
{Dominik} M.,  {Belczynski} K.,  {Fryer} C.,  {Holz} D.~E.,  {Berti} E.,
  {Bulik} T.,  {Mandel} I.,   {O'Shaughnessy} R.,  2013, \mn@doi [\apj]
  {10.1088/0004-637X/779/1/72}, \href
  {http://adsabs.harvard.edu/abs/2013ApJ...779...72D} {779, 72}

\bibitem[\protect\citeauthoryear{{Downing}, {Benacquista}, {Giersz}  \&
  {Spurzem}}{{Downing} et~al.}{2010}]{downing2010}
{Downing} J.~M.~B.,  {Benacquista} M.~J.,  {Giersz} M.,   {Spurzem} R.,  2010,
  \mn@doi [\mnras] {10.1111/j.1365-2966.2010.17040.x}, \href
  {http://adsabs.harvard.edu/abs/2010MNRAS.407.1946D} {407, 1946}

\bibitem[\protect\citeauthoryear{{Drout} et~al.,}{{Drout}
  et~al.}{2017}]{Drout2017}
{Drout} M.~R.,  et~al., 2017, \mn@doi [Science] {10.1126/science.aaq0049},
  \href {https://ui.adsabs.harvard.edu/abs/2017Sci...358.1570D} {358, 1570}

\bibitem[\protect\citeauthoryear{{Dwyer}, {Sigg}, {Ballmer}, {Barsotti},
  {Mavalvala}  \& {Evans}}{{Dwyer} et~al.}{2015}]{Dwyer2015}
{Dwyer} S.,  {Sigg} D.,  {Ballmer} S.~W.,  {Barsotti} L.,  {Mavalvala} N.,
  {Evans} M.,  2015, \mn@doi [\prd] {10.1103/PhysRevD.91.082001}, \href
  {https://ui.adsabs.harvard.edu/abs/2015PhRvD..91h2001D} {91, 082001}

\bibitem[\protect\citeauthoryear{{Eldridge} \& {Stanway}}{{Eldridge} \&
  {Stanway}}{2016}]{eldridge2016}
{Eldridge} J.~J.,  {Stanway} E.~R.,  2016, \mn@doi [\mnras]
  {10.1093/mnras/stw1772}, \href
  {https://ui.adsabs.harvard.edu/abs/2016MNRAS.462.3302E} {462, 3302}

\bibitem[\protect\citeauthoryear{{Fan}, {Zhang}, {Kobayashi}  \&
  {M{\'e}sz{\'a}ros}}{{Fan} et~al.}{2005}]{Fan2005}
{Fan} Y.~Z.,  {Zhang} B.,  {Kobayashi} S.,   {M{\'e}sz{\'a}ros} P.,  2005,
  \mn@doi [\apj] {10.1086/430339}, \href
  {https://ui.adsabs.harvard.edu/abs/2005ApJ...628..867F} {628, 867}

\bibitem[\protect\citeauthoryear{{Fletcher}, {Fermi-GBM Team}  \&
  {GBM-LIGO/Virgo Group}}{{Fletcher} et~al.}{2019}]{FermiGCN190425}
{Fletcher} C.,  {Fermi-GBM Team}  {GBM-LIGO/Virgo Group} 2019, GRB Coordinates
  Network, \href {https://ui.adsabs.harvard.edu/abs/2019GCN.24185....1F}
  {24185, 1}

\bibitem[\protect\citeauthoryear{{Fong} \& {Berger}}{{Fong} \&
  {Berger}}{2013}]{fong2013}
{Fong} W.,  {Berger} E.,  2013, \mn@doi [\apj] {10.1088/0004-637X/776/1/18},
  \href {http://adsabs.harvard.edu/abs/2013ApJ...776...18F} {776, 18}

\bibitem[\protect\citeauthoryear{{Fong}, {Berger}, {Margutti}  \&
  {Zauderer}}{{Fong} et~al.}{2015}]{Fong2015}
{Fong} W.,  {Berger} E.,  {Margutti} R.,   {Zauderer} B.~A.,  2015, \mn@doi
  [\apj] {10.1088/0004-637X/815/2/102}, \href
  {https://ui.adsabs.harvard.edu/abs/2015ApJ...815..102F} {815, 102}

\bibitem[\protect\citeauthoryear{{Foucart}, {Duez}, {Gudinas}, {H{\'e}bert},
  {Kidder}, {Pfeiffer}  \& {Scheel}}{{Foucart} et~al.}{2019}]{Foucart2019}
{Foucart} F.,  {Duez} M.,  {Gudinas} A.,  {H{\'e}bert} F.,  {Kidder} L.,
  {Pfeiffer} H.,   {Scheel} M.,  2019, \mn@doi [\prd]
  {10.1103/PhysRevD.100.104048}, \href
  {https://ui.adsabs.harvard.edu/abs/2019PhRvD.100j4048F} {100, 104048}

\bibitem[\protect\citeauthoryear{{Fragione} \& {Loeb}}{{Fragione} \&
  {Loeb}}{2019}]{fragione2019}
{Fragione} G.,  {Loeb} A.,  2019, \mn@doi [\mnras] {10.1093/mnras/stz1131},
  \href {https://ui.adsabs.harvard.edu/abs/2019MNRAS.486.4443F} {486, 4443}

\bibitem[\protect\citeauthoryear{{Fragione} \& {Silk}}{{Fragione} \&
  {Silk}}{2020}]{fragione2020}
{Fragione} G.,  {Silk} J.,  2020, arXiv e-prints, \href
  {https://ui.adsabs.harvard.edu/abs/2020arXiv200601867F} {p. arXiv:2006.01867}

\bibitem[\protect\citeauthoryear{{Fragione}, {Grishin}, {Leigh}, {Perets}  \&
  {Perna}}{{Fragione} et~al.}{2019a}]{Fragione2019clua}
{Fragione} G.,  {Grishin} E.,  {Leigh} N. W.~C.,  {Perets} H.~B.,   {Perna} R.,
   2019a, \mn@doi [\mnras] {10.1093/mnras/stz1651}, \href
  {https://ui.adsabs.harvard.edu/abs/2019MNRAS.488...47F} {488, 47}

\bibitem[\protect\citeauthoryear{{Fragione}, {Leigh}  \& {Perna}}{{Fragione}
  et~al.}{2019b}]{Fragione2019club}
{Fragione} G.,  {Leigh} N. W.~C.,   {Perna} R.,  2019b, \mn@doi [\mnras]
  {10.1093/mnras/stz1803}, \href
  {https://ui.adsabs.harvard.edu/abs/2019MNRAS.488.2825F} {488, 2825}

\bibitem[\protect\citeauthoryear{{Fryxell} et~al.,}{{Fryxell}
  et~al.}{2000}]{Fryxell2000}
{Fryxell} B.,  et~al., 2000, \mn@doi [\apjs] {10.1086/317361}, \href
  {https://ui.adsabs.harvard.edu/abs/2000ApJS..131..273F} {131, 273}

\bibitem[\protect\citeauthoryear{{Genel}, {Bouch{\'e}}, {Naab}, {Sternberg}  \&
  {Genzel}}{{Genel} et~al.}{2010}]{Genel2010}
{Genel} S.,  {Bouch{\'e}} N.,  {Naab} T.,  {Sternberg} A.,   {Genzel} R.,
  2010, \mn@doi [\apj] {10.1088/0004-637X/719/1/229}, \href
  {https://ui.adsabs.harvard.edu/abs/2010ApJ...719..229G} {719, 229}

\bibitem[\protect\citeauthoryear{{Giacobbo} \& {Mapelli}}{{Giacobbo} \&
  {Mapelli}}{2020}]{giacobbo2020}
{Giacobbo} N.,  {Mapelli} M.,  2020, \mn@doi [\apj] {10.3847/1538-4357/ab7335},
  \href {https://ui.adsabs.harvard.edu/abs/2020ApJ...891..141G} {891, 141}

\bibitem[\protect\citeauthoryear{{Giacobbo}, {Mapelli}  \& {Spera}}{{Giacobbo}
  et~al.}{2018}]{giacobbo2018}
{Giacobbo} N.,  {Mapelli} M.,   {Spera} M.,  2018, \mn@doi [\mnras]
  {10.1093/mnras/stx2933}, \href
  {http://adsabs.harvard.edu/abs/2018MNRAS.474.2959G} {474, 2959}

\bibitem[\protect\citeauthoryear{{Granot}, {Panaitescu}, {Kumar}  \&
  {Woosley}}{{Granot} et~al.}{2002}]{Granot2002}
{Granot} J.,  {Panaitescu} A.,  {Kumar} P.,   {Woosley} S.~E.,  2002, \mn@doi
  [\apjl] {10.1086/340991}, \href
  {https://ui.adsabs.harvard.edu/abs/2002ApJ...570L..61G} {570, L61}

\bibitem[\protect\citeauthoryear{{Guetta} \& {Stella}}{{Guetta} \&
  {Stella}}{2009}]{Guetta2009}
{Guetta} D.,  {Stella} L.,  2009, \mn@doi [\aap] {10.1051/0004-6361:200810493},
  \href {https://ui.adsabs.harvard.edu/abs/2009A&A...498..329G} {498, 329}

\bibitem[\protect\citeauthoryear{{Haggard}, {Nynka}, {Ruan}, {Kalogera},
  {Cenko}, {Evans}  \& {Kennea}}{{Haggard} et~al.}{2017}]{Haggard2017}
{Haggard} D.,  {Nynka} M.,  {Ruan} J.~J.,  {Kalogera} V.,  {Cenko} S.~B.,
  {Evans} P.,   {Kennea} J.~A.,  2017, \mn@doi [\apjl]
  {10.3847/2041-8213/aa8ede}, \href
  {https://ui.adsabs.harvard.edu/abs/2017ApJ...848L..25H} {848, L25}

\bibitem[\protect\citeauthoryear{{Hallinan} et~al.,}{{Hallinan}
  et~al.}{2017}]{Hallinan2017}
{Hallinan} G.,  et~al., 2017, \mn@doi [Science] {10.1126/science.aap9855},
  \href {https://ui.adsabs.harvard.edu/abs/2017Sci...358.1579H} {358, 1579}

\bibitem[\protect\citeauthoryear{{Hobbs}, {Lorimer}, {Lyne}  \&
  {Kramer}}{{Hobbs} et~al.}{2005}]{hobbs2005}
{Hobbs} G.,  {Lorimer} D.~R.,  {Lyne} A.~G.,   {Kramer} M.,  2005, \mn@doi
  [\mnras] {10.1111/j.1365-2966.2005.09087.x}, \href
  {http://adsabs.harvard.edu/abs/2005MNRAS.360..974H} {360, 974}

\bibitem[\protect\citeauthoryear{{Hurley}, {Tout}  \& {Pols}}{{Hurley}
  et~al.}{2002}]{hurley2002}
{Hurley} J.~R.,  {Tout} C.~A.,   {Pols} O.~R.,  2002, \mn@doi [\mnras]
  {10.1046/j.1365-8711.2002.05038.x}, \href
  {http://adsabs.harvard.edu/abs/2002MNRAS.329..897H} {329, 897}

\bibitem[\protect\citeauthoryear{{Kasen}, {Metzger}, {Barnes}, {Quataert}  \&
  {Ramirez-Ruiz}}{{Kasen} et~al.}{2017}]{Kasen2017}
{Kasen} D.,  {Metzger} B.,  {Barnes} J.,  {Quataert} E.,   {Ramirez-Ruiz} E.,
  2017, \mn@doi [\nat] {10.1038/nature24453}, \href
  {https://ui.adsabs.harvard.edu/abs/2017Natur.551...80K} {551, 80}

\bibitem[\protect\citeauthoryear{{Kasliwal} et~al.,}{{Kasliwal}
  et~al.}{2017}]{Kasliwal2017}
{Kasliwal} M.~M.,  et~al., 2017, \mn@doi [Science] {10.1126/science.aap9455},
  \href {https://ui.adsabs.harvard.edu/abs/2017Sci...358.1559K} {358, 1559}

\bibitem[\protect\citeauthoryear{{Kawamura}, {Giacomazzo}, {Kastaun}, {Ciolfi},
  {Endrizzi}, {Baiotti}  \& {Perna}}{{Kawamura} et~al.}{2016}]{Kawamura2016}
{Kawamura} T.,  {Giacomazzo} B.,  {Kastaun} W.,  {Ciolfi} R.,  {Endrizzi} A.,
  {Baiotti} L.,   {Perna} R.,  2016, \mn@doi [\prd]
  {10.1103/PhysRevD.94.064012}, \href
  {https://ui.adsabs.harvard.edu/abs/2016PhRvD..94f4012K} {94, 064012}

\bibitem[\protect\citeauthoryear{{Kiuchi}, {Kyutoku}, {Sekiguchi}  \&
  {Shibata}}{{Kiuchi} et~al.}{2018}]{Kiuchi2018}
{Kiuchi} K.,  {Kyutoku} K.,  {Sekiguchi} Y.,   {Shibata} M.,  2018, \mn@doi
  [\prd] {10.1103/PhysRevD.97.124039}, \href
  {https://ui.adsabs.harvard.edu/abs/2018PhRvD..97l4039K} {97, 124039}

\bibitem[\protect\citeauthoryear{{Kopa{\v{c}}} et~al.,}{{Kopa{\v{c}}}
  et~al.}{2012}]{Kopac2012}
{Kopa{\v{c}}} D.,  et~al., 2012, \mn@doi [\mnras]
  {10.1111/j.1365-2966.2012.21418.x}, \href
  {https://ui.adsabs.harvard.edu/abs/2012MNRAS.424.2392K} {424, 2392}

\bibitem[\protect\citeauthoryear{{Kremer} et~al.,}{{Kremer}
  et~al.}{2020}]{Kremer2020}
{Kremer} K.,  et~al., 2020, \mn@doi [\apjs] {10.3847/1538-4365/ab7919}, \href
  {https://ui.adsabs.harvard.edu/abs/2020ApJS..247...48K} {247, 48}

\bibitem[\protect\citeauthoryear{{Kroupa}}{{Kroupa}}{2001}]{kroupa2001}
{Kroupa} P.,  2001, \mn@doi [\mnras] {10.1046/j.1365-8711.2001.04022.x}, \href
  {http://adsabs.harvard.edu/abs/2001MNRAS.322..231K} {322, 231}

\bibitem[\protect\citeauthoryear{{Kumamoto}, {Fujii}  \& {Tanikawa}}{{Kumamoto}
  et~al.}{2019}]{kumamoto2019}
{Kumamoto} J.,  {Fujii} M.~S.,   {Tanikawa} A.,  2019, \mn@doi [\mnras]
  {10.1093/mnras/stz1068}, \href
  {https://ui.adsabs.harvard.edu/abs/2019MNRAS.486.3942K} {486, 3942}

\bibitem[\protect\citeauthoryear{{Lamb} et~al.,}{{Lamb}
  et~al.}{2019}]{Lamb2019}
{Lamb} G.~P.,  et~al., 2019, \mn@doi [\apjl] {10.3847/2041-8213/aaf96b}, \href
  {https://ui.adsabs.harvard.edu/abs/2019ApJ...870L..15L} {870, L15}

\bibitem[\protect\citeauthoryear{{Lazzati}, {L{\'o}pez-C{\'a}mara},
  {Cantiello}, {Morsony}, {Perna}  \& {Workman}}{{Lazzati}
  et~al.}{2017}]{Lazzati2017}
{Lazzati} D.,  {L{\'o}pez-C{\'a}mara} D.,  {Cantiello} M.,  {Morsony} B.~J.,
  {Perna} R.,   {Workman} J.~C.,  2017, \mn@doi [\apjl]
  {10.3847/2041-8213/aa8f3d}, \href
  {https://ui.adsabs.harvard.edu/abs/2017ApJ...848L...6L} {848, L6}

\bibitem[\protect\citeauthoryear{{Lazzati}, {Perna}, {Morsony}, {Lopez-Camara},
  {Cantiello}, {Ciolfi}, {Giacomazzo}  \& {Workman}}{{Lazzati}
  et~al.}{2018}]{Lazzati2018}
{Lazzati} D.,  {Perna} R.,  {Morsony} B.~J.,  {Lopez-Camara} D.,  {Cantiello}
  M.,  {Ciolfi} R.,  {Giacomazzo} B.,   {Workman} J.~C.,  2018, \mn@doi [\prl]
  {10.1103/PhysRevLett.120.241103}, \href
  {https://ui.adsabs.harvard.edu/abs/2018PhRvL.120x1103L} {120, 241103}

\bibitem[\protect\citeauthoryear{{Lyman} et~al.,}{{Lyman}
  et~al.}{2018}]{Lyman2018}
{Lyman} J.~D.,  et~al., 2018, \mn@doi [Nature Astronomy]
  {10.1038/s41550-018-0511-3}, \href
  {https://ui.adsabs.harvard.edu/abs/2018NatAs...2..751L} {2, 751}

\bibitem[\protect\citeauthoryear{{Mandel} \& {Broekgaarden}}{{Mandel} \&
  {Broekgaarden}}{2021}]{mandel2021}
{Mandel} I.,  {Broekgaarden} F.~S.,  2021, arXiv e-prints, \href
  {https://ui.adsabs.harvard.edu/abs/2021arXiv210714239M} {p. arXiv:2107.14239}

\bibitem[\protect\citeauthoryear{{Mandhai}, {Lamb}, {Tanvir}, {Bray}, {Nixon},
  {Eyles-Ferris}, {Levan}  \& {Gompertz}}{{Mandhai} et~al.}{2021}]{Mandhai2021}
{Mandhai} S.,  {Lamb} G.~P.,  {Tanvir} N.~R.,  {Bray} J.,  {Nixon} C.~J.,
  {Eyles-Ferris} R.~A.~J.,  {Levan} A.~J.,   {Gompertz} B.~P.,  2021, arXiv
  e-prints, \href {https://ui.adsabs.harvard.edu/abs/2021arXiv210909714M} {p.
  arXiv:2109.09714}

\bibitem[\protect\citeauthoryear{{Mapelli}}{{Mapelli}}{2016}]{mapelli2016}
{Mapelli} M.,  2016, \mn@doi [\mnras] {10.1093/mnras/stw869}, \href
  {http://adsabs.harvard.edu/abs/2016MNRAS.459.3432M} {459, 3432}

\bibitem[\protect\citeauthoryear{{Mapelli} \& {Giacobbo}}{{Mapelli} \&
  {Giacobbo}}{2018}]{Mapelli2018}
{Mapelli} M.,  {Giacobbo} N.,  2018, \mn@doi [\mnras] {10.1093/mnras/sty1613},
  \href {http://adsabs.harvard.edu/abs/2018MNRAS.479.4391M} {479, 4391}

\bibitem[\protect\citeauthoryear{{Mapelli}, {Giacobbo}, {Ripamonti}  \&
  {Spera}}{{Mapelli} et~al.}{2017}]{mapelli2017}
{Mapelli} M.,  {Giacobbo} N.,  {Ripamonti} E.,   {Spera} M.,  2017, \mn@doi
  [\mnras] {10.1093/mnras/stx2123}, \href
  {http://adsabs.harvard.edu/abs/2017MNRAS.472.2422M} {472, 2422}

\bibitem[\protect\citeauthoryear{{Mapelli}, {Giacobbo}, {Toffano}, {Ripamonti},
  {Bressan}, {Spera}  \& {Branchesi}}{{Mapelli} et~al.}{2018}]{Mapelli2018b}
{Mapelli} M.,  {Giacobbo} N.,  {Toffano} M.,  {Ripamonti} E.,  {Bressan} A.,
  {Spera} M.,   {Branchesi} M.,  2018, \mn@doi [\mnras]
  {10.1093/mnras/sty2663}, \href
  {http://adsabs.harvard.edu/abs/2018MNRAS.481.5324M} {481, 5324}

\bibitem[\protect\citeauthoryear{{Mapelli}, {Giacobbo}, {Santoliquido}  \&
  {Artale}}{{Mapelli} et~al.}{2019}]{Mapelli2019}
{Mapelli} M.,  {Giacobbo} N.,  {Santoliquido} F.,   {Artale} M.~C.,  2019,
  \mn@doi [\mnras] {10.1093/mnras/stz1150}, \href
  {https://ui.adsabs.harvard.edu/abs/2019MNRAS.487....2M} {487, 2}

\bibitem[\protect\citeauthoryear{{Mapelli} et~al.,}{{Mapelli}
  et~al.}{2021}]{Mapelli2021a}
{Mapelli} M.,  et~al., 2021, \mn@doi [\mnras] {10.1093/mnras/stab1334}, \href
  {https://ui.adsabs.harvard.edu/abs/2021MNRAS.505..339M} {505, 339}

\bibitem[\protect\citeauthoryear{{Marchant}, {Langer}, {Podsiadlowski},
  {Tauris}  \& {Moriya}}{{Marchant} et~al.}{2016}]{marchant2016}
{Marchant} P.,  {Langer} N.,  {Podsiadlowski} P.,  {Tauris} T.~M.,   {Moriya}
  T.~J.,  2016, \mn@doi [\aap] {10.1051/0004-6361/201628133}, \href
  {http://adsabs.harvard.edu/abs/2016A%26A...588A..50M} {588, A50}

\bibitem[\protect\citeauthoryear{{Margutti} et~al.,}{{Margutti}
  et~al.}{2012}]{Margutti2012}
{Margutti} R.,  et~al., 2012, \mn@doi [\apj] {10.1088/0004-637X/756/1/63},
  \href {https://ui.adsabs.harvard.edu/abs/2012ApJ...756...63M} {756, 63}

\bibitem[\protect\citeauthoryear{{Margutti} et~al.,}{{Margutti}
  et~al.}{2017}]{Margutti2017}
{Margutti} R.,  et~al., 2017, \mn@doi [\apjl] {10.3847/2041-8213/aa9057}, \href
  {https://ui.adsabs.harvard.edu/abs/2017ApJ...848L..20M} {848, L20}

\bibitem[\protect\citeauthoryear{{Margutti} et~al.,}{{Margutti}
  et~al.}{2018}]{Margutti2018}
{Margutti} R.,  et~al., 2018, \mn@doi [\apjl] {10.3847/2041-8213/aab2ad}, \href
  {https://ui.adsabs.harvard.edu/abs/2018ApJ...856L..18M} {856, L18}

\bibitem[\protect\citeauthoryear{{McKernan}, {Ford}, {Lyra}  \&
  {Perets}}{{McKernan} et~al.}{2012}]{mckernan2012}
{McKernan} B.,  {Ford} K.~E.~S.,  {Lyra} W.,   {Perets} H.~B.,  2012, \mn@doi
  [\mnras] {10.1111/j.1365-2966.2012.21486.x}, \href
  {https://ui.adsabs.harvard.edu/abs/2012MNRAS.425..460M} {425, 460}

\bibitem[\protect\citeauthoryear{{McKernan} et~al.,}{{McKernan}
  et~al.}{2018}]{mckernan2018}
{McKernan} B.,  et~al., 2018, \mn@doi [\apj] {10.3847/1538-4357/aadae5}, \href
  {https://ui.adsabs.harvard.edu/abs/2018ApJ...866...66M} {866, 66}

\bibitem[\protect\citeauthoryear{{Miller} \& {Lauburg}}{{Miller} \&
  {Lauburg}}{2009}]{miller2009}
{Miller} M.~C.,  {Lauburg} V.~M.,  2009, \mn@doi [\apj]
  {10.1088/0004-637X/692/1/917}, \href
  {http://adsabs.harvard.edu/abs/2009ApJ...692..917M} {692, 917}

\bibitem[\protect\citeauthoryear{{Moffett} et~al.,}{{Moffett}
  et~al.}{2016}]{Moffett2016}
{Moffett} A.~J.,  et~al., 2016, \mn@doi [\mnras] {10.1093/mnras/stw1861}, \href
  {https://ui.adsabs.harvard.edu/abs/2016MNRAS.462.4336M} {462, 4336}

\bibitem[\protect\citeauthoryear{{Mooley} et~al.,}{{Mooley}
  et~al.}{2018}]{Mooley2018}
{Mooley} K.~P.,  et~al., 2018, \mn@doi [\nat] {10.1038/s41586-018-0486-3},
  \href {https://ui.adsabs.harvard.edu/abs/2018Natur.561..355M} {561, 355}

\bibitem[\protect\citeauthoryear{{Most}, {Papenfort}  \& {Rezzolla}}{{Most}
  et~al.}{2019}]{Most2019}
{Most} E.~R.,  {Papenfort} L.~J.,   {Rezzolla} L.,  2019, \mn@doi [\mnras]
  {10.1093/mnras/stz2809}, \href
  {https://ui.adsabs.harvard.edu/abs/2019MNRAS.490.3588M} {490, 3588}

\bibitem[\protect\citeauthoryear{{Murguia-Berthier} et~al.,}{{Murguia-Berthier}
  et~al.}{2021}]{Murguia2021}
{Murguia-Berthier} A.,  et~al., 2021, \mn@doi [\apj]
  {10.3847/1538-4357/ac1119}, \href
  {https://ui.adsabs.harvard.edu/abs/2021ApJ...919...95M} {919, 95}

\bibitem[\protect\citeauthoryear{{Nakar}, {Gal-Yam}  \& {Fox}}{{Nakar}
  et~al.}{2006}]{Nakar2006}
{Nakar} E.,  {Gal-Yam} A.,   {Fox} D.~B.,  2006, \mn@doi [\apj]
  {10.1086/505855}, \href
  {https://ui.adsabs.harvard.edu/abs/2006ApJ...650..281N} {650, 281}

\bibitem[\protect\citeauthoryear{{Neijssel} et~al.,}{{Neijssel}
  et~al.}{2019}]{neijssel2019}
{Neijssel} C.~J.,  et~al., 2019, \mn@doi [\mnras] {10.1093/mnras/stz2840},
  \href {https://ui.adsabs.harvard.edu/abs/2019MNRAS.490.3740N} {490, 3740}

\bibitem[\protect\citeauthoryear{{Nissanke}, {Kasliwal}  \&
  {Georgieva}}{{Nissanke} et~al.}{2013}]{Nissanke2013}
{Nissanke} S.,  {Kasliwal} M.,   {Georgieva} A.,  2013, \mn@doi [\apj]
  {10.1088/0004-637X/767/2/124}, \href
  {https://ui.adsabs.harvard.edu/abs/2013ApJ...767..124N} {767, 124}

\bibitem[\protect\citeauthoryear{{O'Leary}, {Rasio}, {Fregeau}, {Ivanova}  \&
  {O'Shaughnessy}}{{O'Leary} et~al.}{2006}]{oleary2006}
{O'Leary} R.~M.,  {Rasio} F.~A.,  {Fregeau} J.~M.,  {Ivanova} N.,
  {O'Shaughnessy} R.,  2006, \mn@doi [\apj] {10.1086/498446}, \href
  {http://adsabs.harvard.edu/abs/2006ApJ...637..937O} {637, 937}

\bibitem[\protect\citeauthoryear{{O'Shaughnessy}, {Belczynski}  \&
  {Kalogera}}{{O'Shaughnessy} et~al.}{2008}]{Oshaughnessy2008}
{O'Shaughnessy} R.,  {Belczynski} K.,   {Kalogera} V.,  2008, \mn@doi [\apj]
  {10.1086/526334}, \href
  {https://ui.adsabs.harvard.edu/abs/2008ApJ...675..566O} {675, 566}

\bibitem[\protect\citeauthoryear{{O'Shaughnessy}, {Kalogera}  \&
  {Belczynski}}{{O'Shaughnessy} et~al.}{2010}]{oshaughnessy2010}
{O'Shaughnessy} R.,  {Kalogera} V.,   {Belczynski} K.,  2010, \mn@doi [\apj]
  {10.1088/0004-637X/716/1/615}, \href
  {https://ui.adsabs.harvard.edu/abs/2010ApJ...716..615O} {716, 615}

\bibitem[\protect\citeauthoryear{{Palmese} et~al.,}{{Palmese}
  et~al.}{2017}]{Palmese2017}
{Palmese} A.,  et~al., 2017, \mn@doi [\apjl] {10.3847/2041-8213/aa9660}, \href
  {https://ui.adsabs.harvard.edu/abs/2017ApJ...849L..34P} {849, L34}

\bibitem[\protect\citeauthoryear{{Panaitescu} \& {Kumar}}{{Panaitescu} \&
  {Kumar}}{2000}]{Panaitescu2000}
{Panaitescu} A.,  {Kumar} P.,  2000, \mn@doi [\apj] {10.1086/317090}, \href
  {https://ui.adsabs.harvard.edu/abs/2000ApJ...543...66P} {543, 66}

\bibitem[\protect\citeauthoryear{{Paterson} et~al.,}{{Paterson}
  et~al.}{2021}]{Paterson2021}
{Paterson} K.,  et~al., 2021, \mn@doi [\apj] {10.3847/1538-4357/abeb71}, \href
  {https://ui.adsabs.harvard.edu/abs/2021ApJ...912..128P} {912, 128}

\bibitem[\protect\citeauthoryear{{Perna} \& {Belczynski}}{{Perna} \&
  {Belczynski}}{2002}]{Perna2002}
{Perna} R.,  {Belczynski} K.,  2002, \mn@doi [\apj] {10.1086/339571}, \href
  {https://ui.adsabs.harvard.edu/abs/2002ApJ...570..252P} {570, 252}

\bibitem[\protect\citeauthoryear{{Perna}, {Wang}, {Farr}, {Leigh}  \&
  {Cantiello}}{{Perna} et~al.}{2019}]{Perna2019}
{Perna} R.,  {Wang} Y.-H.,  {Farr} W.~M.,  {Leigh} N.,   {Cantiello} M.,  2019,
  \mn@doi [\apjl] {10.3847/2041-8213/ab2336}, \href
  {https://ui.adsabs.harvard.edu/abs/2019ApJ...878L...1P} {878, L1}

\bibitem[\protect\citeauthoryear{{Perna}, {Lazzati}  \& {Cantiello}}{{Perna}
  et~al.}{2021a}]{Perna2021a}
{Perna} R.,  {Lazzati} D.,   {Cantiello} M.,  2021a, \mn@doi [\apjl]
  {10.3847/2041-8213/abd319}, \href
  {https://ui.adsabs.harvard.edu/abs/2021ApJ...906L...7P} {906, L7}

\bibitem[\protect\citeauthoryear{{Perna}, {Tagawa}, {Haiman}  \&
  {Bartos}}{{Perna} et~al.}{2021b}]{Perna2021b}
{Perna} R.,  {Tagawa} H.,  {Haiman} Z.,   {Bartos} I.,  2021b, \mn@doi [\apj]
  {10.3847/1538-4357/abfdb4}, \href
  {https://ui.adsabs.harvard.edu/abs/2021ApJ...915...10P} {915, 10}

\bibitem[\protect\citeauthoryear{{Pian} et~al.,}{{Pian}
  et~al.}{2017}]{pian2017}
{Pian} E.,  et~al., 2017, \mn@doi [\nat] {10.1038/nature24298}, \href
  {http://adsabs.harvard.edu/abs/2017Natur.551...67P} {551, 67}

\bibitem[\protect\citeauthoryear{{Pillepich} et~al.,}{{Pillepich}
  et~al.}{2017}]{pillepich2017}
{Pillepich} A.,  et~al., 2017, preprint, \href
  {http://adsabs.harvard.edu/abs/2017arXiv170302970P} {} (\mn@eprint {arXiv}
  {1703.02970})

\bibitem[\protect\citeauthoryear{{Pillepich} et~al.,}{{Pillepich}
  et~al.}{2019}]{Pillepich2019}
{Pillepich} A.,  et~al., 2019, \mn@doi [\mnras] {10.1093/mnras/stz2338}, \href
  {https://ui.adsabs.harvard.edu/abs/2019MNRAS.490.3196P} {490, 3196}

\bibitem[\protect\citeauthoryear{{Piranomonte} et~al.,}{{Piranomonte}
  et~al.}{2008}]{Piranomonte2008}
{Piranomonte} S.,  et~al., 2008, \mn@doi [\aap] {10.1051/0004-6361:200810547},
  \href {https://ui.adsabs.harvard.edu/abs/2008A&A...491..183P} {491, 183}

\bibitem[\protect\citeauthoryear{{Piro} et~al.,}{{Piro}
  et~al.}{2019}]{Piro2019}
{Piro} L.,  et~al., 2019, \mn@doi [\mnras] {10.1093/mnras/sty3047}, \href
  {https://ui.adsabs.harvard.edu/abs/2019MNRAS.483.1912P} {483, 1912}

\bibitem[\protect\citeauthoryear{{Podsiadlowski}, {Langer}, {Poelarends},
  {Rappaport}, {Heger}  \& {Pfahl}}{{Podsiadlowski}
  et~al.}{2004}]{podsiadlowski2004}
{Podsiadlowski} P.,  {Langer} N.,  {Poelarends} A.~J.~T.,  {Rappaport} S.,
  {Heger} A.,   {Pfahl} E.,  2004, \mn@doi [\apj] {10.1086/421713}, \href
  {http://adsabs.harvard.edu/abs/2004ApJ...612.1044P} {612, 1044}

\bibitem[\protect\citeauthoryear{{Portegies Zwart} \& {McMillan}}{{Portegies
  Zwart} \& {McMillan}}{2000}]{portegieszwart2000}
{Portegies Zwart} S.~F.,  {McMillan} S.~L.~W.,  2000, \mn@doi [\apjl]
  {10.1086/312422}, \href {http://adsabs.harvard.edu/abs/2000ApJ...528L..17P}
  {528, L17}

\bibitem[\protect\citeauthoryear{{Portegies Zwart} \& {Yungelson}}{{Portegies
  Zwart} \& {Yungelson}}{1998}]{portegieszwart1998}
{Portegies Zwart} S.~F.,  {Yungelson} L.~R.,  1998, \aap, \href
  {http://adsabs.harvard.edu/abs/1998A%26A...332..173P} {332, 173}

\bibitem[\protect\citeauthoryear{{Punturo} et~al.,}{{Punturo}
  et~al.}{2010}]{Punturo2010}
{Punturo} M.,  et~al., 2010, \mn@doi [Classical and Quantum Gravity]
  {10.1088/0264-9381/27/19/194002}, \href
  {https://ui.adsabs.harvard.edu/abs/2010CQGra..27s4002P} {27, 194002}

\bibitem[\protect\citeauthoryear{{Radice}}{{Radice}}{2017}]{Radice2017}
{Radice} D.,  2017, \mn@doi [\apjl] {10.3847/2041-8213/aa6483}, \href
  {https://ui.adsabs.harvard.edu/abs/2017ApJ...838L...2R} {838, L2}

\bibitem[\protect\citeauthoryear{{Rasskazov} \& {Kocsis}}{{Rasskazov} \&
  {Kocsis}}{2019}]{rasskazov2019}
{Rasskazov} A.,  {Kocsis} B.,  2019, \mn@doi [\apj] {10.3847/1538-4357/ab2c74},
  \href {https://ui.adsabs.harvard.edu/abs/2019ApJ...881...20R} {881, 20}

\bibitem[\protect\citeauthoryear{{Rastello}, {Mapelli}, {Di Carlo}, {Giacobbo},
  {Santoliquido}, {Spera}  \& {Ballone}}{{Rastello}
  et~al.}{2020}]{rastello2020}
{Rastello} S.,  {Mapelli} M.,  {Di Carlo} U.~N.,  {Giacobbo} N.,
  {Santoliquido} F.,  {Spera} M.,   {Ballone} A.,  2020, arXiv e-prints, \href
  {https://ui.adsabs.harvard.edu/abs/2020arXiv200302277R} {p. arXiv:2003.02277}

\bibitem[\protect\citeauthoryear{{Resmi} et~al.,}{{Resmi}
  et~al.}{2018}]{Resmi2018}
{Resmi} L.,  et~al., 2018, \mn@doi [\apj] {10.3847/1538-4357/aae1a6}, \href
  {https://ui.adsabs.harvard.edu/abs/2018ApJ...867...57R} {867, 57}

\bibitem[\protect\citeauthoryear{{Rodriguez} \& {Loeb}}{{Rodriguez} \&
  {Loeb}}{2018}]{rodriguez2018}
{Rodriguez} C.~L.,  {Loeb} A.,  2018, \mn@doi [\apjl]
  {10.3847/2041-8213/aae377}, \href
  {http://adsabs.harvard.edu/abs/2018ApJ...866L...5R} {866, L5}

\bibitem[\protect\citeauthoryear{{Rodriguez-Gomez} et~al.,}{{Rodriguez-Gomez}
  et~al.}{2015}]{RodriguezGomez2015}
{Rodriguez-Gomez} V.,  et~al., 2015, \mn@doi [\mnras] {10.1093/mnras/stv264},
  \href {https://ui.adsabs.harvard.edu/abs/2015MNRAS.449...49R} {449, 49}

\bibitem[\protect\citeauthoryear{{Rodriguez}, {Morscher}, {Pattabiraman},
  {Chatterjee}, {Haster}  \& {Rasio}}{{Rodriguez} et~al.}{2015}]{rodriguez2015}
{Rodriguez} C.~L.,  {Morscher} M.,  {Pattabiraman} B.,  {Chatterjee} S.,
  {Haster} C.-J.,   {Rasio} F.~A.,  2015, \mn@doi [Physical Review Letters]
  {10.1103/PhysRevLett.115.051101}, \href
  {http://adsabs.harvard.edu/abs/2015PhRvL.115e1101R} {115, 051101}

\bibitem[\protect\citeauthoryear{{Rodriguez}, {Chatterjee}  \&
  {Rasio}}{{Rodriguez} et~al.}{2016}]{rodriguez2016}
{Rodriguez} C.~L.,  {Chatterjee} S.,   {Rasio} F.~A.,  2016, \mn@doi [\prd]
  {10.1103/PhysRevD.93.084029}, \href
  {http://adsabs.harvard.edu/abs/2016PhRvD..93h4029R} {93, 084029}

\bibitem[\protect\citeauthoryear{{Rose}, {Torrey}, {Lee}  \& {Bartos}}{{Rose}
  et~al.}{2021}]{Rose2021}
{Rose} J.~C.,  {Torrey} P.,  {Lee} K.~H.,   {Bartos} I.,  2021, \mn@doi [\apj]
  {10.3847/1538-4357/abe405}, \href
  {https://ui.adsabs.harvard.edu/abs/2021ApJ...909..207R} {909, 207}

\bibitem[\protect\citeauthoryear{{Rossi}, {Lazzati}, {Salmonson}  \&
  {Ghisellini}}{{Rossi} et~al.}{2004}]{Rossi2004}
{Rossi} E.~M.,  {Lazzati} D.,  {Salmonson} J.~D.,   {Ghisellini} G.,  2004,
  \mn@doi [\mnras] {10.1111/j.1365-2966.2004.08165.x}, \href
  {https://ui.adsabs.harvard.edu/abs/2004MNRAS.354...86R} {354, 86}

\bibitem[\protect\citeauthoryear{{Ruan}, {Nynka}, {Haggard}, {Kalogera}  \&
  {Evans}}{{Ruan} et~al.}{2018}]{Ruan2018}
{Ruan} J.~J.,  {Nynka} M.,  {Haggard} D.,  {Kalogera} V.,   {Evans} P.,  2018,
  \mn@doi [\apjl] {10.3847/2041-8213/aaa4f3}, \href
  {https://ui.adsabs.harvard.edu/abs/2018ApJ...853L...4R} {853, L4}

\bibitem[\protect\citeauthoryear{{Ruiz}, {Tsokaros}  \& {Shapiro}}{{Ruiz}
  et~al.}{2020}]{Ruiz2020}
{Ruiz} M.,  {Tsokaros} A.,   {Shapiro} S.~L.,  2020, \mn@doi [\prd]
  {10.1103/PhysRevD.101.064042}, \href
  {https://ui.adsabs.harvard.edu/abs/2020PhRvD.101f4042R} {101, 064042}

\bibitem[\protect\citeauthoryear{{Sakamoto} et~al.,}{{Sakamoto}
  et~al.}{2019}]{SwiftGCN190425}
{Sakamoto} T.,  et~al., 2019, GRB Coordinates Network, \href
  {https://ui.adsabs.harvard.edu/abs/2019GCN.24184....1S} {24184, 1}

\bibitem[\protect\citeauthoryear{{Salafia}, {Ghirlanda}, {Ascenzi}  \&
  {Ghisellini}}{{Salafia} et~al.}{2019}]{Salafia2019}
{Salafia} O.~S.,  {Ghirlanda} G.,  {Ascenzi} S.,   {Ghisellini} G.,  2019,
  \mn@doi [\aap] {10.1051/0004-6361/201935831}, \href
  {https://ui.adsabs.harvard.edu/abs/2019A&A...628A..18S} {628, A18}

\bibitem[\protect\citeauthoryear{{Saleem}, {Resmi}, {Misra}, {Pai}  \&
  {Arun}}{{Saleem} et~al.}{2018}]{Saleem2018a}
{Saleem} M.,  {Resmi} L.,  {Misra} K.,  {Pai} A.,   {Arun} K.~G.,  2018,
  \mn@doi [\mnras] {10.1093/mnras/stx3104}, \href
  {https://ui.adsabs.harvard.edu/abs/2018MNRAS.474.5340S} {474, 5340}

\bibitem[\protect\citeauthoryear{{Samsing}}{{Samsing}}{2018}]{samsing2018}
{Samsing} J.,  2018, \mn@doi [\prd] {10.1103/PhysRevD.97.103014}, \href
  {http://adsabs.harvard.edu/abs/2018PhRvD..97j3014S} {97, 103014}

\bibitem[\protect\citeauthoryear{{Samsing}, {MacLeod}  \&
  {Ramirez-Ruiz}}{{Samsing} et~al.}{2014}]{samsing2014}
{Samsing} J.,  {MacLeod} M.,   {Ramirez-Ruiz} E.,  2014, \mn@doi [\apj]
  {10.1088/0004-637X/784/1/71}, \href
  {https://ui.adsabs.harvard.edu/abs/2014ApJ...784...71S} {784, 71}

\bibitem[\protect\citeauthoryear{{Sana} et~al.,}{{Sana}
  et~al.}{2012}]{sana2012}
{Sana} H.,  et~al., 2012, \mn@doi [Science] {10.1126/science.1223344}, \href
  {http://adsabs.harvard.edu/abs/2012Sci...337..444S} {337, 444}

\bibitem[\protect\citeauthoryear{{Santoliquido}, {Mapelli}, {Bouffanais},
  {Giacobbo}, {Di Carlo}, {Rastello}, {Artale}  \& {Ballone}}{{Santoliquido}
  et~al.}{2020}]{santoliquido2020}
{Santoliquido} F.,  {Mapelli} M.,  {Bouffanais} Y.,  {Giacobbo} N.,  {Di Carlo}
  U.~N.,  {Rastello} S.,  {Artale} M.~C.,   {Ballone} A.,  2020, arXiv
  e-prints, \href {https://ui.adsabs.harvard.edu/abs/2020arXiv200409533S} {p.
  arXiv:2004.09533}

\bibitem[\protect\citeauthoryear{{Santoliquido}, {Mapelli}, {Giacobbo},
  {Bouffanais}  \& {Artale}}{{Santoliquido} et~al.}{2021}]{Santoliquido2021}
{Santoliquido} F.,  {Mapelli} M.,  {Giacobbo} N.,  {Bouffanais} Y.,   {Artale}
  M.~C.,  2021, \mn@doi [\mnras] {10.1093/mnras/stab280}, \href
  {https://ui.adsabs.harvard.edu/abs/2021MNRAS.502.4877S} {502, 4877}

\bibitem[\protect\citeauthoryear{{Sari}, {Piran}  \& {Narayan}}{{Sari}
  et~al.}{1998}]{Sari1998}
{Sari} R.,  {Piran} T.,   {Narayan} R.,  1998, \mn@doi [\apjl]
  {10.1086/311269}, \href
  {https://ui.adsabs.harvard.edu/abs/1998ApJ...497L..17S} {497, L17}

\bibitem[\protect\citeauthoryear{{Schaye} et~al.,}{{Schaye}
  et~al.}{2015}]{Schaye2015}
{Schaye} J.,  et~al., 2015, \mn@doi [\mnras] {10.1093/mnras/stu2058}, \href
  {https://ui.adsabs.harvard.edu/abs/2015MNRAS.446..521S} {446, 521}

\bibitem[\protect\citeauthoryear{{Schutz}}{{Schutz}}{2011}]{Schutz2011}
{Schutz} B.~F.,  2011, \mn@doi [Classical and Quantum Gravity]
  {10.1088/0264-9381/28/12/125023}, \href
  {https://ui.adsabs.harvard.edu/abs/2011CQGra..28l5023S} {28, 125023}

\bibitem[\protect\citeauthoryear{{Smartt} et~al.,}{{Smartt}
  et~al.}{2017}]{Smartt2017}
{Smartt} S.~J.,  et~al., 2017, \mn@doi [\nat] {10.1038/nature24303}, \href
  {https://ui.adsabs.harvard.edu/abs/2017Natur.551...75S} {551, 75}

\bibitem[\protect\citeauthoryear{Soares-Santos et~al.,}{Soares-Santos
  et~al.}{2017}]{soares-santos2017}
Soares-Santos M.,  et~al., 2017, The Astrophysical Journal Letters, 848, L16

\bibitem[\protect\citeauthoryear{{Song}, {Ai}, {Wang}, {Xing}, {Gao}  \&
  {Zhang}}{{Song} et~al.}{2019}]{Song2019}
{Song} H.-R.,  {Ai} S.-K.,  {Wang} M.-H.,  {Xing} N.,  {Gao} H.,   {Zhang} B.,
  2019, \mn@doi [\apjl] {10.3847/2041-8213/ab3921}, \href
  {https://ui.adsabs.harvard.edu/abs/2019ApJ...881L..40S} {881, L40}

\bibitem[\protect\citeauthoryear{{Spera}, {Mapelli}, {Giacobbo}, {Trani},
  {Bressan}  \& {Costa}}{{Spera} et~al.}{2019}]{spera2019}
{Spera} M.,  {Mapelli} M.,  {Giacobbo} N.,  {Trani} A.~A.,  {Bressan} A.,
  {Costa} G.,  2019, \mn@doi [\mnras] {10.1093/mnras/stz359}, \href
  {https://ui.adsabs.harvard.edu/abs/2019MNRAS.485..889S} {485, 889}

\bibitem[\protect\citeauthoryear{{Stevenson}, {Berry}  \& {Mandel}}{{Stevenson}
  et~al.}{2017}]{stevenson2017}
{Stevenson} S.,  {Berry} C.~P.~L.,   {Mandel} I.,  2017, preprint, \href
  {http://adsabs.harvard.edu/abs/2017arXiv170306873S} {} (\mn@eprint {arXiv}
  {1703.06873})

\bibitem[\protect\citeauthoryear{{Stone}, {Metzger}  \& {Haiman}}{{Stone}
  et~al.}{2017}]{stone2017}
{Stone} N.~C.,  {Metzger} B.~D.,   {Haiman} Z.,  2017, \mn@doi [\mnras]
  {10.1093/mnras/stw2260}, \href
  {https://ui.adsabs.harvard.edu/abs/2017MNRAS.464..946S} {464, 946}

\bibitem[\protect\citeauthoryear{{Tagawa}, {Haiman}  \& {Kocsis}}{{Tagawa}
  et~al.}{2020a}]{Tagawa2020a}
{Tagawa} H.,  {Haiman} Z.,   {Kocsis} B.,  2020a, \mn@doi [\apj]
  {10.3847/1538-4357/ab7922}, \href
  {https://ui.adsabs.harvard.edu/abs/2020ApJ...892...36T} {892, 36}

\bibitem[\protect\citeauthoryear{{Tagawa}, {Haiman}  \& {Kocsis}}{{Tagawa}
  et~al.}{2020b}]{Tagawa2020b}
{Tagawa} H.,  {Haiman} Z.,   {Kocsis} B.,  2020b, \mn@doi [\apj]
  {10.3847/1538-4357/ab9b8c}, \href
  {https://ui.adsabs.harvard.edu/abs/2020ApJ...898...25T} {898, 25}

\bibitem[\protect\citeauthoryear{{Tagawa}, {Kocsis}, {Haiman}, {Bartos},
  {Omukai}  \& {Samsing}}{{Tagawa} et~al.}{2021a}]{Tagawa2021a}
{Tagawa} H.,  {Kocsis} B.,  {Haiman} Z.,  {Bartos} I.,  {Omukai} K.,
  {Samsing} J.,  2021a, \mn@doi [\apjl] {10.3847/2041-8213/abd4d3}, \href
  {https://ui.adsabs.harvard.edu/abs/2021ApJ...907L..20T} {907, L20}

\bibitem[\protect\citeauthoryear{{Tagawa}, {Kocsis}, {Haiman}, {Bartos},
  {Omukai}  \& {Samsing}}{{Tagawa} et~al.}{2021b}]{Tagawa2021b}
{Tagawa} H.,  {Kocsis} B.,  {Haiman} Z.,  {Bartos} I.,  {Omukai} K.,
  {Samsing} J.,  2021b, \mn@doi [\apj] {10.3847/1538-4357/abd555}, \href
  {https://ui.adsabs.harvard.edu/abs/2021ApJ...908..194T} {908, 194}

\bibitem[\protect\citeauthoryear{{Tang}, {Eldridge}, {Stanway}  \&
  {Bray}}{{Tang} et~al.}{2020}]{tang2019}
{Tang} P.~N.,  {Eldridge} J.~J.,  {Stanway} E.~R.,   {Bray} J.~C.,  2020,
  \mn@doi [\mnras] {10.1093/mnrasl/slz183}, \href
  {https://ui.adsabs.harvard.edu/abs/2020MNRAS.493L...6T} {493, L6}

\bibitem[\protect\citeauthoryear{{Tanikawa}, {Susa}, {Yoshida}, {Trani}  \&
  {Kinugawa}}{{Tanikawa} et~al.}{2020}]{tanikawa2020}
{Tanikawa} A.,  {Susa} H.,  {Yoshida} T.,  {Trani} A.~A.,   {Kinugawa} T.,
  2020, arXiv e-prints, \href
  {https://ui.adsabs.harvard.edu/abs/2020arXiv200801890T} {p. arXiv:2008.01890}

\bibitem[\protect\citeauthoryear{{Tanvir} et~al.,}{{Tanvir}
  et~al.}{2017}]{Tanvir2017}
{Tanvir} N.~R.,  et~al., 2017, \mn@doi [\apjl] {10.3847/2041-8213/aa90b6},
  \href {https://ui.adsabs.harvard.edu/abs/2017ApJ...848L..27T} {848, L27}

\bibitem[\protect\citeauthoryear{{Troja} et~al.,}{{Troja}
  et~al.}{2017}]{Troja2017}
{Troja} E.,  et~al., 2017, \mn@doi [\nat] {10.1038/nature24290}, \href
  {https://ui.adsabs.harvard.edu/abs/2017Natur.551...71T} {551, 71}

\bibitem[\protect\citeauthoryear{{Tutukov} \& {Yungelson}}{{Tutukov} \&
  {Yungelson}}{1973}]{tutukov1973}
{Tutukov} A.,  {Yungelson} L.,  1973, Nauchnye Informatsii, \href
  {http://adsabs.harvard.edu/abs/1973NInfo..27...70T} {27, 70}

\bibitem[\protect\citeauthoryear{{Vigna-G{\'o}mez} et~al.,}{{Vigna-G{\'o}mez}
  et~al.}{2018}]{vignagomez2018}
{Vigna-G{\'o}mez} A.,  et~al., 2018, \mn@doi [\mnras] {10.1093/mnras/sty2463},
  \href {https://ui.adsabs.harvard.edu/abs/2018MNRAS.481.4009V} {481, 4009}

\bibitem[\protect\citeauthoryear{{Vogelsberger}, {Genel}, {Sijacki}, {Torrey},
  {Springel}  \& {Hernquist}}{{Vogelsberger} et~al.}{2013}]{vogelsberger2013}
{Vogelsberger} M.,  {Genel} S.,  {Sijacki} D.,  {Torrey} P.,  {Springel} V.,
  {Hernquist} L.,  2013, \mn@doi [\mnras] {10.1093/mnras/stt1789}, \href
  {http://adsabs.harvard.edu/abs/2013MNRAS.436.3031V} {436, 3031}

\bibitem[\protect\citeauthoryear{{Voss} \& {Tauris}}{{Voss} \&
  {Tauris}}{2003}]{voss2003}
{Voss} R.,  {Tauris} T.~M.,  2003, \mn@doi [\mnras]
  {10.1046/j.1365-8711.2003.06616.x}, \href
  {http://adsabs.harvard.edu/abs/2003MNRAS.342.1169V} {342, 1169}

\bibitem[\protect\citeauthoryear{{Wu} \& {MacFadyen}}{{Wu} \&
  {MacFadyen}}{2019}]{Wu2019}
{Wu} Y.,  {MacFadyen} A.,  2019, \mn@doi [\apjl] {10.3847/2041-8213/ab2fd4},
  \href {https://ui.adsabs.harvard.edu/abs/2019ApJ...880L..23W} {880, L23}

\bibitem[\protect\citeauthoryear{{Yang} et~al.,}{{Yang}
  et~al.}{2019a}]{Yang2019b}
{Yang} Y.,  et~al., 2019a, \mn@doi [\prl] {10.1103/PhysRevLett.123.181101},
  \href {https://ui.adsabs.harvard.edu/abs/2019PhRvL.123r1101Y} {123, 181101}

\bibitem[\protect\citeauthoryear{{Yang}, {Bartos}, {Haiman}, {Kocsis},
  {M{\'a}rka}, {Stone}  \& {M{\'a}rka}}{{Yang} et~al.}{2019b}]{yang2019}
{Yang} Y.,  {Bartos} I.,  {Haiman} Z.,  {Kocsis} B.,  {M{\'a}rka} Z.,  {Stone}
  N.~C.,   {M{\'a}rka} S.,  2019b, \mn@doi [\apj] {10.3847/1538-4357/ab16e3},
  \href {https://ui.adsabs.harvard.edu/abs/2019ApJ...876..122Y} {876, 122}

\bibitem[\protect\citeauthoryear{{Yu} et~al.,}{{Yu} et~al.}{2021}]{Yu2021}
{Yu} J.,  et~al., 2021, \mn@doi [\apj] {10.3847/1538-4357/ac0628}, \href
  {https://ui.adsabs.harvard.edu/abs/2021ApJ...916...54Y} {916, 54}

\bibitem[\protect\citeauthoryear{{Zevin}, {Samsing}, {Rodriguez}, {Haster}  \&
  {Ramirez-Ruiz}}{{Zevin} et~al.}{2019}]{zevin2019}
{Zevin} M.,  {Samsing} J.,  {Rodriguez} C.,  {Haster} C.-J.,   {Ramirez-Ruiz}
  E.,  2019, \mn@doi [\apj] {10.3847/1538-4357/aaf6ec}, \href
  {https://ui.adsabs.harvard.edu/abs/2019ApJ...871...91Z} {871, 91}

\bibitem[\protect\citeauthoryear{{Zhu}, {Zhang}, {Yu}  \& {Gao}}{{Zhu}
  et~al.}{2021}]{Zhu2021}
{Zhu} J.-P.,  {Zhang} B.,  {Yu} Y.-W.,   {Gao} H.,  2021, \mn@doi [\apjl]
  {10.3847/2041-8213/abd412}, \href
  {https://ui.adsabs.harvard.edu/abs/2021ApJ...906L..11Z} {906, L11}

\bibitem[\protect\citeauthoryear{{Ziosi}, {Mapelli}, {Branchesi}  \&
  {Tormen}}{{Ziosi} et~al.}{2014}]{ziosi2014}
{Ziosi} B.~M.,  {Mapelli} M.,  {Branchesi} M.,   {Tormen} G.,  2014, \mn@doi
  [\mnras] {10.1093/mnras/stu824}, \href
  {http://adsabs.harvard.edu/abs/2014MNRAS.441.3703Z} {441, 3703}

\bibitem[\protect\citeauthoryear{{de Mink} \& {Mandel}}{{de Mink} \&
  {Mandel}}{2016}]{demink2016}
{de Mink} S.~E.,  {Mandel} I.,  2016, \mn@doi [\mnras] {10.1093/mnras/stw1219},
  \href {http://adsabs.harvard.edu/abs/2016MNRAS.460.3545D} {460, 3545}

\makeatother
\end{thebibliography}


\bsp	
\label{lastpage}
\end{document}